\documentclass[conference]{IEEEtran}

\usepackage{mathtools,dsfont}
\usepackage{psfrag,bbm,graphicx}
\usepackage{algorithm,algorithmic}
\usepackage{comment,balance}
\usepackage{epstopdf}
\usepackage{epsfig}
\usepackage{multicol}
\usepackage{multirow, cite}
\usepackage{amsfonts}

\newtheorem{corollary}{Corollary}

\newtheorem{lemma}{Lemma}
\newtheorem{remark}{Remark}

\newtheorem{theorem}{Theorem}
\newtheorem{definition}{Definition}

\newtheorem{assumption}{Assumption}

\def\EE{{\mathbb{E}}}
\def\PP{{\mathbb{P}}}

\def \OO {\mathrm{O}}
\def \oo {\mathrm{o}}

\newfont{\mycrnotice}{ptmr8t at 7pt}
\newfont{\myconfname}{ptmri8t at 7pt}

\begin{document}

\title{Content Replication in Large Distributed Caches}
\author{Sharayu Moharir and Nikhil Karamchandani \\
	Department of Electrical Engineering \\
	Indian Institute of Technology, Bombay \\
	Email: sharayum@ee.iitb.ac.in, nikhilk@ee.iitb.ac.in
}
\maketitle
\thispagestyle{empty}

\begin{abstract}
	In this paper, we consider the algorithmic task of content replication and request routing in a distributed caching system consisting of a central server and a large number of caches, each with limited storage and service capabilities. We study a time-slotted system where in each time-slot, a large batch of requests has to be matched to a large number of caches, where each request can be served by any cache which stores the requested content. All requests which cannot be served by the caches are served by fetching the requested content from the central server. The goal is to minimize the transmission rate from the central server. 
	
	We use a novel mapping between our content replication problem and the Knapsack problem to prove a lower bound on the transmission rate for any content replication policy. Using insights obtained from the mapping, we propose a content replication policy - \emph{Knapsack Storage} - which achieves this lower bound. While it intuitively makes sense to replicate the more popular contents on a larger number of caches, surprisingly, in certain cases, the Knapsack Storage policy chooses not to replicate the most popular contents on the caches at all. 
	
	
\end{abstract}


\section{introduction}

Video on Demand (VoD) services like Netflix \cite{netflix} and Youtube \cite{Youtube} account for ever-increasing fractions of Internet traffic. This fraction is expected to cross $50\%$ by 2018 \cite{Cisco}. Most VoD services use distributed Content Delivery Networks (CDNs) to serve their customers. Caching content closer to the network edge, i.e., close to the end users is an effective mechanism to reduce the load on the network backbone, thus reducing the bandwidth consumption of the network. The scope for cost savings and performance benefits from caching increases as the cost of memory continues to drop at a higher rate than that of transmission gear \cite{borst2010distributed}.

\begin{figure}[t]
	\begin{center}
		\includegraphics[scale=0.3]{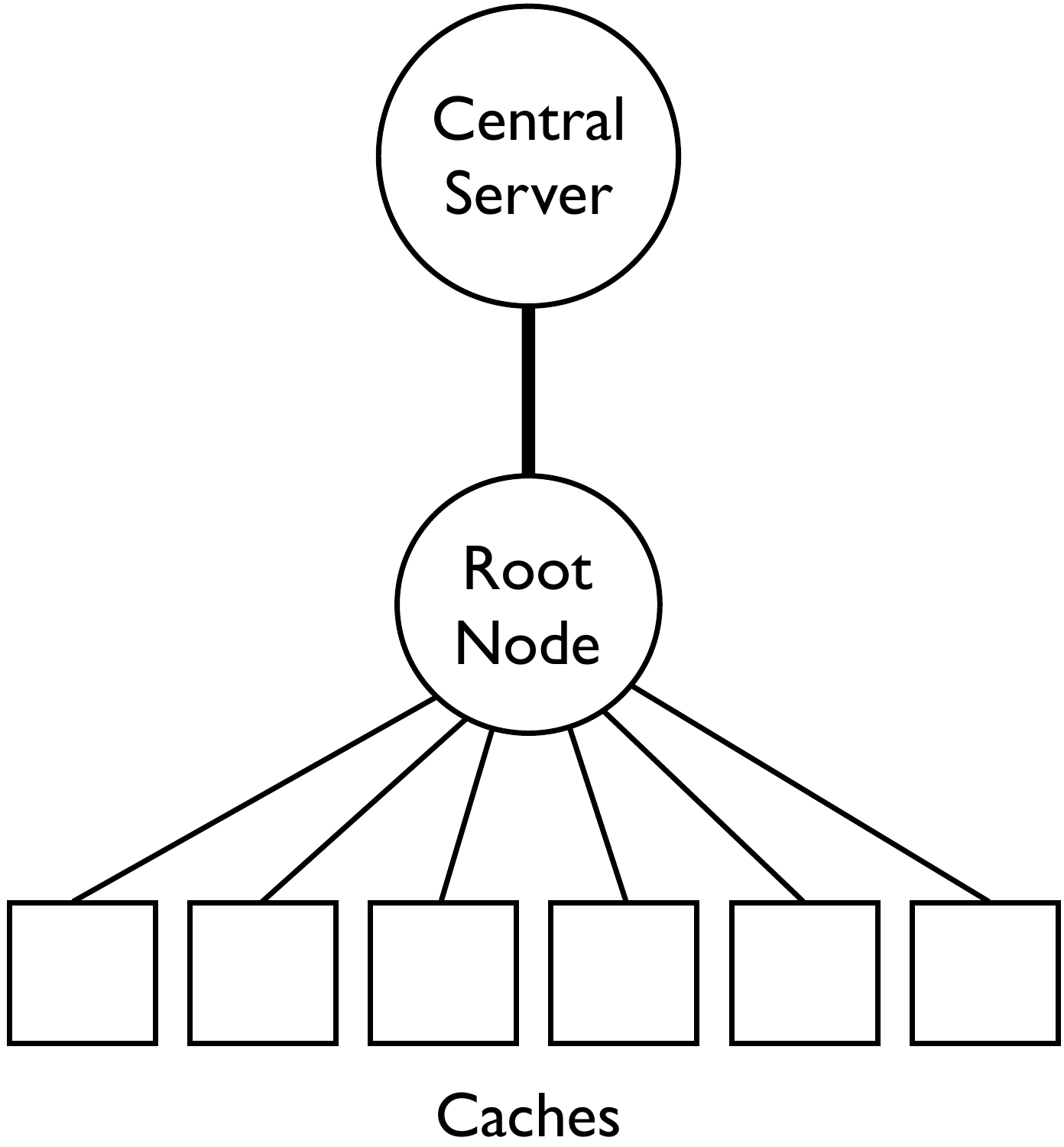}
		\caption{\sl Cache Cluster. \label{fig:cache_cluster}}
	\end{center}
\end{figure}

In this work, we focus on the task of content replication in a cache cluster consisting of multiple caches, each with limited storage and service capabilities, connected to a common root node that has a link to the central server as shown in Figure \ref{fig:cache_cluster}. For example, the root node could represent an ISP using multiple caches to serve users in a specific geographical area and central server represents core network. This cache cluster need not necessarily be a stand-alone network, but could in fact be a part of a larger tree topology \cite{borst2010distributed}.

As discussed in \cite{maddah2014fundamental, pedarsani2014online, niesen2014coded, hachem2014multi, zhang2015coded, hachem2014content }, the system operates in two distinct phases. The first phase is the \emph{placement phase} where content is replicated on the caches based on the statistics of the user demands. In the second phase, called the \emph{delivery phase}, user requests are served using the caches and transmission from the central server via the root node. Since each cache can only store a small subset of the contents offered by the VoD service, optimizing content replication in the placement phase is critical for the efficient service in the delivery phase.

Most popular VoD services serve a large number of users and have extremely large content catalogs, e.g., YouTube serves over a billion users and offers close to a billion hours of content \cite{YoutubeStatistics}. Both these factors need to be incorporated into algorithm design and performance analysis for such networks. Motivated by this, we focus on the asymptotic setting where the number of different contents offered by the service is large, and a large batch of user requests arrive at the beginning of each time-slot. Since the caches have limited service capacity, the incoming requests have to be matched to the caches such that no cache is allocated more than one request in each time-slot. All requests which can't be served by the local caches are served by the central server which stores the entire catalog of contents offered by the VoD service. The goal is to optimize content replication on the caches and design a computationally efficient request matching policy to minimize the rate of transmission from the central server to the root node. 

Recently, two other settings have been used to model such CDNs. Similar to our setting, the setting in \cite{LLM12, LLM13, SGSS14, SGSS14_2} allows each request to be served by any one of the caches, as long as no cache serves more than one request at any given time. However, in \cite{LLM12, LLM13, SGSS14, SGSS14_2}, the central server communicates with each user via an independent link. The setting in \cite{maddah2014fundamental, pedarsani2014online, niesen2014coded, hachem2014multi, zhang2015coded, hachem2014content } is different from our setting as it pre-assigns each user to a specific cache. 

We compare and contrast the optimal content replication strategies for our setting and the two settings described above. Our larger goal and a key contribution of this work is to develop a comprehensive understanding of the effects of the structure of the underlying network on the nature of the optimal content replication strategies. 

\subsection{Contributions}
The main contributions of our work can be summarized as follows. 

\begin{enumerate}
	\item[(i)] \emph{Converse Results:} We use a novel mapping between our content replication problem and the Knapsack problem to lower bound the expected transmission rate for all content replication policies which do not use coded caching (Theorem \ref{theorem:converse}). We also prove a stronger information-theoretic converse which lower bounds the expected transmission rate for all content replication policies if content popularity follows the Zipf's law (Theorem \ref{theorem:converse_information_theoretic}). These results can be extended to the case where content popularity follows the Zipf-Mandelbrot law \cite{melendi2009multivariate}.
	
	\item[(ii)] \emph{Our Policy:} Using the insights obtained from the mapping, we propose a content replication and request matching policy which solves a fractional knapsack problem to determine which contents to store on the caches. We show that our policy is optimal if content popularity follows the Zipf's law (Theorem \ref{theorem:performance:our_policy}). This result can be extended to the case where content popularity follows the Zipf-Mandelbrot law. 
	
	Surprisingly, in certain cases, our content replication policy does not cache the most popular contents. Intuitively, in order to serve all the requests for a popular content via the caches, the content needs to be replicated on a large number of caches, since each cache can only serve one request at a time. It follows that, at times, it is better to serve all the requests  for a popular content via a single transmission from the central server, instead of replicating it on a large number of caches, thus using up a lot of memory resources.  
	
	\item[(iii)] \emph{Comparison with other Settings: } As mentioned before, for the settings studied in \cite{maddah2014fundamental, maddah2013decentralized, zhang2015coded}, coded caching is necessary for optimal performance. The only difference between our setting and the setting in \cite{maddah2014fundamental, maddah2013decentralized, zhang2015coded} is that in \cite{maddah2014fundamental, maddah2013decentralized, zhang2015coded}, each request is pre-assigned to a specific cache, whereas, in our setting, each request can be matched to any one of the caches. We show that this added flexibility in matching requests to caches eliminates the need of coding for optimal performance. 
	Another key insight we obtain is that unlike our policy, the optimal policies for the two alternative settings discussed above, store the more popular contents on a larger number of caches.  
\end{enumerate}

%
%

\section{Setting}
\label{section:system_model}
We study a distributed caching system consisting of a central server, and $m$ caches, each with limited storage and service capabilities, connected to the central server via a root node (Figure \ref{fig:cache_cluster}). The system offers a content catalog consisting of $n$ contents, where $n$ scales linearly with respect to $m$, i.e., $n = cm$ where $c$ is a constant. We are interested in the asymptotic performance of the system as $n,m \rightarrow \infty$ to model the enormous content catalogs offered by most large scale content delivery systems.

\subsection{Storage Model}
The central server stores the entire catalog of contents offered by the content delivery system, and each of the $m$ caches can store $k$ units of data. Let $b_i$ denote the units of storage required to store a copy of Content $i$. 

\subsection{Request Model}
We consider a time-slotted system where a batch of $\tilde{m} = \Theta(m)$ requests arrives at the beginning of each time-slot. Each request is generated according to an independent and identically distributed process. The probability that a request is for Content $i$ is denoted by $p_i$. We use the Zipf distribution to model the popularity of contents as empirical studies of many VoD services have shown that the content popularity profile matches well with such distributions, see e.g., \cite{liu2013measurement,liu2012server,BC99,YZ06,IRF04,VA02,fricker2012impact}. 
\begin{assumption}
	\label{assumption:Zipf_popularity}
	Zipf Popularity \\
	Without loss of generality, contents are indexed in decreasing order of popularity with $p_i \propto i^{-\beta}$, where $\beta>0$ is a constant, known as the Zipf parameter. 
\end{assumption}
All the results in this paper can be extended to the case where content popularity follows the Zipf-Mandelbrot law \cite{melendi2009multivariate}.

\subsection{Service Model}
Each request can be served in one time-slot by any one cache which stores the requested content, and each cache can serve at most one request in each time-slot. In each time-slot, all the requests that cannot be served by the caches are served by fetching the requested content from the central server via the root node.


\subsection{Goal}
The goal is to determine what to store in the caches, and how to match incoming requests to the caches in order to minimize the expected transmission rate from the central server needed to serve all the requests arriving in a time-slot.

\section{Preliminaries}
\label{section:preliminaries}
The fractional knapsack problem \cite{goodrich2006algorithm} can informally be defined as follows: choose items to keep in the knapsack such that the cumulative value of the items is maximized, while ensuring that the cumulative weight of the items is not more than the knapsack's capacity. Formally, if the total capacity of the knapsack is $W$, item $j$ has value $v_j$ and weight $w_j$, the fractional knapsack problem is defined as:
\begin{eqnarray*}
	&& \max \displaystyle \sum_{j=1}^J x_j v_j \\
	\text{s.t. } &&\displaystyle \sum_{j=1}^J x_j w_j \leq W, \\
	&& 0 \leq x_j \leq 1, \text{ } \forall j.
\end{eqnarray*}
WLOG, let the items be indexed in decreasing order of value to weight ratio, i.e.,
$
\frac{v_1}{w_1} \geq \frac{v_2}{w_2} \geq ... \geq \frac{v_J}{w_J}. 
$

\noindent Let $j^*$ be such that
\begin{eqnarray*}
	\sum_{j=1}^{j^*-1} w_j \leq W, \text{ and } \sum_{j=1}^{j^*} w_j > W. 
\end{eqnarray*}
The solution to the fractional knapsack problem is:
\begin{eqnarray*}
	x_j = \begin{cases}
		1, & \text{for } j < j^*,  \\
		\dfrac{W - \sum_{j=1}^{j^*-1} w_j}{w_{j^*}}, & \text{for } j = j^*,\\
		0 & \text{otherwise}.
	\end{cases}
\end{eqnarray*}
\begin{remark}
	The solution to the fractional knapsack problem can be computed in $\OO(J \log J)$ time. 
\end{remark}

\section{Main Results and Discussion}
\label{section:main_results}

In this section, we state and discuss our main results. The proofs are provided in Section \ref{section:proofs} and the Appendix. 


\subsection{Converse without Coding}
In this section, we compute a lower bound on the expected transmission rate from the central server for all policies which do not use coding. Theorem \ref{theorem:converse} provides a lower bound on the expected tranmission rate for the case where each request arrives according to an independent and identically distributed process.
\begin{theorem}
	\label{theorem:converse}
	Consider a distributed cache system with $n$ contents, $m$ caches, and a batch of $\tilde{m}$ requests arriving at the beginning of each time-slot. Each request is generated according to an independent and identically distributed process, and the probability that a request is for Content $i$ is denoted by $p_i$. Let $R^*_{\text{NC}}$ denote the minimum transmission rate required to serve all requests arriving in a time-slot, and let Content $i$ need $b_i$ units of storage. Then, we have that,
	\begin{eqnarray*}
		\EE[R^*_{\text{NC}}] &\geq& 
		\displaystyle \sum_{i=1}^n b_i (1-(1-p_i)^{\tilde{m}}) - \text{O}^*, \\
		\text{where, O}^* &=& \max \displaystyle \sum_{i=1}^n \displaystyle \sum_{u=1}^{b_i} x_{i,u} (1-(1-p_i)^{\tilde{m}}) \\
		&& \text{s.t. }  \displaystyle \sum_{i=1}^n \displaystyle \sum_{u=1}^{b_i} x_{i,u} \max\{\tilde{m}p_i,1 \}  \leq mk, \\
		&& 0 \leq x_{i,u} \leq 1, \text{ } \forall i,u.
	\end{eqnarray*}
\end{theorem}

\begin{remark} The quantity $\text{O}^*$ defined in Theorem \ref{theorem:converse} is the solution to the fractional knapsack problem described in Section \ref{section:preliminaries} with:
	\begin{itemize}
		\item[--] The value of bit $u$ of Content $i$, $$v_{i,u}=1-(1-p_i)^{\tilde{m}},$$ is the probability that Content $i$ is requested at least once in the time-slot. 
		\item[--] The weight of bit $u$ of Content $i$, $$w_{i,u} = \lceil \tilde{m}p_i \rceil,$$ where $\tilde{m}p_i$ is the expected number of requests for Content $i$ in a time-slot. 
		\item[--] The capacity of the knapsack, $$W = mk,$$ is the total memory of the $m$ caches. 
	\end{itemize}
	$x_{i,u}=1$ implies that $\lceil \tilde{m}p_i \rceil$ copies of bit $u$ of Content $i$ are stored in the knapsack, and, $x_{i,u}=0$ implies that bit $u$ of Content $i$ is not stored in the knapsack. Theorem \ref{theorem:converse} lower bounds the expected transmission rate by 
	$$\displaystyle \sum_{i=1}^n \sum_{u=1}^{b_i} (1-x_{i,u})(1-(1-p_i)^{\tilde{m}}),$$
	which is the expected number of bits not stored in knapsack and requested at least once.  
\end{remark}
Next, we evaluate the lower bound on the expected transmission rate obtained in Theorem  \ref{theorem:converse} for the case where content popularity follows the Zipf distribution. Numerous empirical studies for many content delivery systems have shown that the distribution of popularities matches well with such distributions, see e.g., \cite{liu2013measurement,liu2012server,BC99,YZ06,IRF04,VA02,fricker2012impact}. 
\begin{corollary}
	\label{corollary:converse_zipf}
	Consider a distributed cache system with $n$ contents, $m$ caches, and a batch of $m$ requests arriving at the beginning of each time-slot. Each request is generated according to an independent and identically distributed process, and the probability that a request is for Content $i$ is denoted by $p_i$.
	If the $p_i$s follow the Zipf distribution with Zipf parameter $\beta$, i.e., $p_i \propto i^{-\beta}$,
	\begin{eqnarray*}
		\EE[R^*_{\text{NC}}] &\geq& \displaystyle \sum_{i=1}^n b_i \bigg(1-\bigg(1-\frac{p_1}{i^{\beta}}\bigg)^m\bigg) - \text{O}^*_{\text{Zipf}}, \\
		\text{where, O}^*_{\text{Zipf}} &=& \max  \displaystyle \sum_{i=1}^n \displaystyle \sum_{u=1}^{b_i} x_{i,u} \bigg(1-\bigg(1-\frac{p_1}{i^{\beta}}\bigg)^m\bigg) \\
		&& \text{s.t. }  \displaystyle \sum_{i=1}^n \displaystyle \sum_{u=1}^{b_i} x_{i,u} \max\bigg\{\tilde{m}\frac{p_1}{i^{\beta}},1 \bigg\}  \leq mk, \\
		&& 0 \leq x_{i,v} \leq 1, \text{ } \forall i,u,
	\end{eqnarray*}
	and $p_1 = \bigg(\displaystyle \sum_{i=1}^n i^{-\beta}\bigg)^{-1}$.
\end{corollary}


Let $\tilde{i} = (mp_1)^{1/\beta}$ and $r_i$ be the value to weight ratio of a bit of content $i$. We have that, 
\begin{eqnarray*}
	r_i = \dfrac{v_i}{w_i} = \begin{cases}
		\dfrac{1-(1-p_i)^m}{mp_i}, & \text{ for } i \leq \tilde{i}-1, \\
		1-(1-p_i)^m, & \text{   for } i \geq \tilde{i}.
	\end{cases}
\end{eqnarray*}
Given this,
\begin{eqnarray*}
	\frac{\text{d}r_i}{\text{d}i} >0,  \text{ for } i \leq \tilde{i}-1, \text{ and }\frac{\text{d}r_i}{\text{d}i} <0,  \text{ for } i \geq \tilde{i}.
\end{eqnarray*}

\begin{figure}
	\begin{center}
		\includegraphics[width=3.25 in]{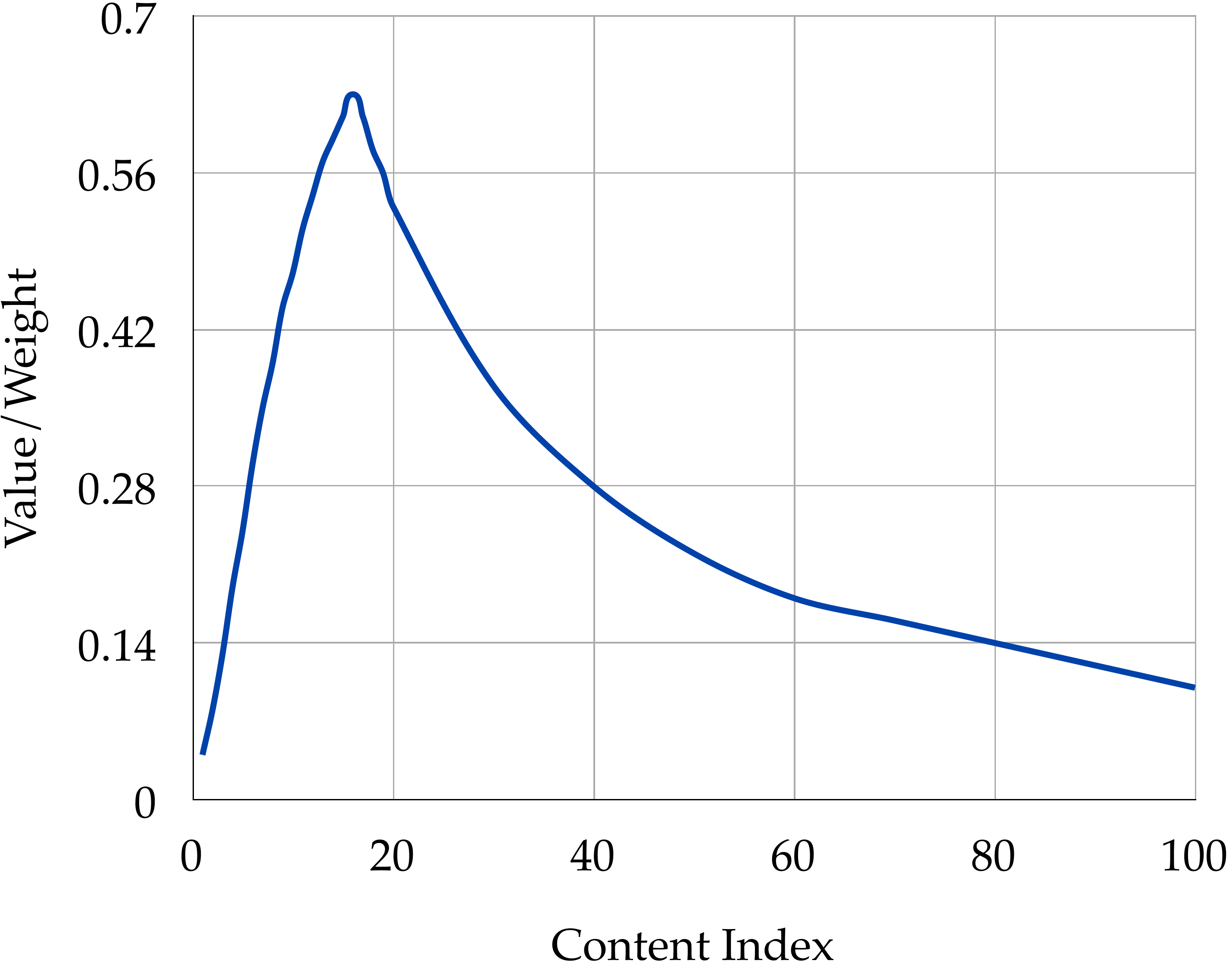}
		\caption{Value to weight ratio when content popularity follows the Zipf distribution for $n=m =100$, and $\beta = 1.2$).}\label{Zipf_ratio}
	\end{center}
\end{figure}

Therefore, $r_i$ increases from $i=1$ to $\tilde{i}-1$ and decreases from $i=\tilde{i}$ to $n$. For example, Figure \ref{Zipf_ratio} illustrates how the ratio of the value to weight ratio for $n=m =100$, and $\beta = 1.2$ varies as a function of content index. 

Given this, and the fact that  the solution to the fractional knapsack problem (Section \ref{section:preliminaries}) is obtained by ranking items in decreasing order of value/weight ratio and choosing the maximum number of highest ranked items such that their cumulative weight is less that the knapsack capacity, the optimal solution to the fractional knapsack solution has the following structure: $\exists$ $i_{\text{min}}$, $i_{\text{max}}$ with $i_{\text{min}} \leq \tilde{i} \leq i_{\text{max}}$, such that, 
\begin{eqnarray*}
	x_{i,v} = \begin{cases}
		1, & \forall v, \text{ if } i_{\text{min}} \leq \tilde{i} \leq i_{\text{max}}-1, \\
		1, & \text{for some values of $v$ if } i =i_{\text{min}} \text{ and } i_{\text{max}}, \\
		0, & \text{otherwise}.
	\end{cases}
\end{eqnarray*}

We optimize over $i_{\text{min}}$ and $i_{\text{max}}$ to get a lower bound on the expected transmission rate for particular values of $c$, $k$ and $m$. For example, Table \ref{table:Zipf_ratio} shows the results for the case where $b_i = b$ $\forall i$, and content popularity follows the Zipf distribution with Zipf Parameter $\beta$, such that $1 < \beta <2$. Typical values of $\beta$ for most Video on Demand services lie between 0.6 and 2 \cite{liu2013measurement,liu2012server,BC99,YZ06,IRF04,VA02,fricker2012impact}. For $\beta<1$, the Zipf distribution is not well defined in the limit as $n \rightarrow \infty$, and therefore, for the asymptotic results obtained in Table \ref{table:Zipf_ratio}, we focus our attention on the case where $1 < \beta <2$.

\begin{remark} 
	\label{remark:what_to_store}
	The key insights from the optimal solution to the fractional knapsack problem in Corollary \ref{corollary:converse_zipf} are: 
	\begin{enumerate}
		\item[--] For contents expected to be requested at least once, i.e., Contents $i$ such that $mp_i \geq 1$, it is optimal to store contents with lower popularity. Intuitively, given that two contents are going to be requested at least once each, all the requests for the less popular content can be served using fewer caches and a lesser amount of storage than the more popular content. Therefore, between the two contents, storing the less popular content reduces the transmission rate by $b$ units using fewer memory resources. 
		\item[--] For contents expected to be requested at most once, i.e., Contents $i$ such that $mp_i < 1$, it is optimal to store the more popular contents. Intuitively, between two contents with the same weight, storing the more popular content increases the probability of reducing the transmission rate required to serve incoming requests, while using the same amount of memory resources.
	\end{enumerate}
\end{remark}

\begin{table}
	\begin{center}
		\begin{tabular}{| l| c|}
			\hline
			& $\EE[R^*_{\text{NC}}]$ \\ 
			\hline
			Case 1: $ c-k/b = \Omega(1)$ & $\Omega(n^{2-\beta})$ \\
			Case 2: $ c-k/b = \Theta(n^{-\epsilon})$, $0 <\epsilon \leq \tilde{\epsilon}$ & $\Omega(n^{2-\beta - \epsilon})$ \\
			Case 3: $ |k/b-c| = \OO(n^{-\tilde{\epsilon}})$, & $\Omega\big(n^{\frac{2-\beta}{\beta}}\big)$ \\
			Case 4: $ k/b-c = \Theta(n^{-\epsilon})$, $0 <\epsilon \leq \tilde{\epsilon}$ & $\Omega(n^{\frac{\epsilon}{\beta-1}})$ \\
			Case 5: $k/b-c \geq 1 - \oo(n^{1/\beta})$ & $0$ \\
			\hline
		\end{tabular}
	\end{center}
	\caption{Converse results for the case where content popularity follows the Zipf distribution with Zipf Parameter $\beta$, such that $1 < \beta <2$. Here, $\tilde{\epsilon} = (2-\beta)(\beta-1)/\beta$, $k$ is the storage per cache, and $c = \frac{\text{Number of contents}}{\text{Number of caches}}$. }\label{table:Zipf_ratio}
\end{table}

%

\subsection{Our Policy}
\subsubsection{Storage Policy: Knapsack Storage}
We use the insights obtained from Theorem \ref{theorem:converse} to design a storage policy. In Theorem \ref{theorem:converse}, we use the solution to the fractional knapsack problem with appropriate values and weights for the contents to lower bound the expected transmission rate. Inspired by this, we propose a storage policy called the Knapsack Storage policy. We describe the Knapsack Storage policy in two parts. \\

\noindent \textbf{Knapsack Storage: Part 1} -- The first part of the Knapsack Storage policy determines how many caches each content is stored on by solving a fractional Knapsack problem. The parameters of the fractional knapsack problem are as follows:
\begin{itemize}
	\item[--] The value of Content $i$, $$v_{i}=1-(1-p_i)^m,$$ is the probability that Content $i$ is requested at least once in the time-slot. 	
	\item[--] The weight of Content $i$ ($w_i$) represents the number of caches Content $i$ will be stored on if selected by the Knapsack problem. If we decide to store a content on the caches, we would like to ensure that all requests for that content can be served by the caches, so that the content need not be transmitted by the central server. To ensure this, we fix $w_i$ to be high enough to ensure that with high probability, i.e., with probability $\rightarrow 1$ as $\tilde{m},m,n \rightarrow \infty$, the number of requests for Content $i$ in a time-slot is $\leq w_i$. We use the following values for the $w_is$:
	\begin{eqnarray*}
		w_i = \begin{cases}
			m, & \text{if } p_i = p^* = \displaystyle \max_{j} p_j, \\
			\big\lceil \big(1 + \frac{p^*}{2}\big) \tilde{m}p_i \big\rceil, & \text{if } p^* > p_i \geq \frac{(\log m)^2}{\tilde{m}}, \\
			\big\lceil 4 p^* (\log m)^2 \big\rceil, & \text{if } \frac{(\log m)^2}{\tilde{m}} > p_i \geq \frac{1}{\tilde{m}(\log m)^2}, \\
			1, & \text{if } \frac{1}{\tilde{m}(\log m)^2} > p_i > 0. 
		\end{cases} 
	\end{eqnarray*}
	\item[--] For simpilicity, we assume that all contents are of equal size and need $b$ units of storage each.
\end{itemize}
Figure \ref{fig:knapsack_storage_Part 1} formally describes Knapsack Storage: Part 1.

\begin{figure}[h]
	\hrule
	\vspace{0.1in}
	\begin{algorithmic}[1]
		\STATE Solve the following fractional knapsack problem
		\begin{eqnarray*}
			\max &&  \displaystyle \sum_{i=1}^n x_i v_i \\
			\text{s.t. }  &&b \displaystyle \sum_{i=1}^n x_i w_i  \leq mk, \\
			&& 0 \leq x_i \leq 1, \text{ } \forall i,
		\end{eqnarray*}
		where $v_i$s and $w_i$s are as defined in Definition \ref{def:knapsack_zipf}.
		\STATE The set of contents to be stored $$S = \{\lfloor x_i \rfloor w_i \text{ copies of Content } i, 1 \leq i \leq n\}.$$
	\end{algorithmic}
	\vspace{0.1in}
	\hrule
	\caption{Knapsack Storage: Part 1 -- \sl Determines how many caches each content is stored on.}
	\label{fig:knapsack_storage_Part 1}
\end{figure}

\begin{remark} Knapsack Storage: Part 1 can be implemented in $\OO(n \log n)$ time, where $n$ is the number of contents in the content catalog. 
	
	Since the value of Content $i$, $v_{i}=1-(1-p_i)^m,$ is the probability that Content $i$ is requested at least once in the time-slot, $n - \sum_{i=1}^n x_{i} v_i,$
	is the expected number of contents that are not stored in the knapsack and requested at least once. As a result, maximizing $\sum_{i=1}^n x_{i} v_i$ minimizes the expected number of contents that are not stored in the knapsack and requested at least once, which is equivalent to minimizing the expected transmission rate. 
	
	Recall from Section \ref{section:preliminaries} that the optimal solution to the fractional knapsack problem prioritizes selecting contents with larger value to weight ratios. Therefore, for certain values of the system parameters ($n$, $m$, $\tilde{m}$, $k$), the optimal solution to the fractional knapsack problem in Figure \ref{fig:knapsack_storage_Part 1} does not store the most popular contents on the caches. As discussed in Remark \ref{remark:what_to_store}, intuitively, in order to serve all the requests for a popular content via the caches, the content needs to be replicated on a large number of caches, since each cache can only serve one request at a time. It follows that, at times, it is better to serve all the requests for a popular content via a single transmission from the central server, instead of replicating it on a large number of caches, thus using up a lot of memory resources.  
	
\end{remark}

\textbf{Knapsack Storage: Part 2} -- The next decision to be made is which contents to store on which caches, i.e., how to partition the set of contents selected by Knapsack Storage: Part 1 (Figure \ref{fig:knapsack_storage_Part 1}) into $m$ groups with $k/b$ contents each.

\begin{figure}[h]
	\hrule
	\vspace{0.1in}
	\begin{algorithmic}[1]
		\STATE Order content copies in $S$ obtained in Knapsack Storage: Part 1 (Figure \ref{fig:knapsack_storage_Part 1}) in increasing order of content index.   
		\STATE Store content copy ranked $r$ in the ordered sequence on cache $((r-1)\mod m + 1)$.
	\end{algorithmic}
	\vspace{0.1in}
	\hrule
	\caption{Knapsack Storage: Part 2 -- \sl Determines which contents to store on each cache.}
	\label{fig:knapsack_storage_Part 2}
\end{figure}
\begin{figure}
	\begin{center}
		\includegraphics[width=1.8 in]{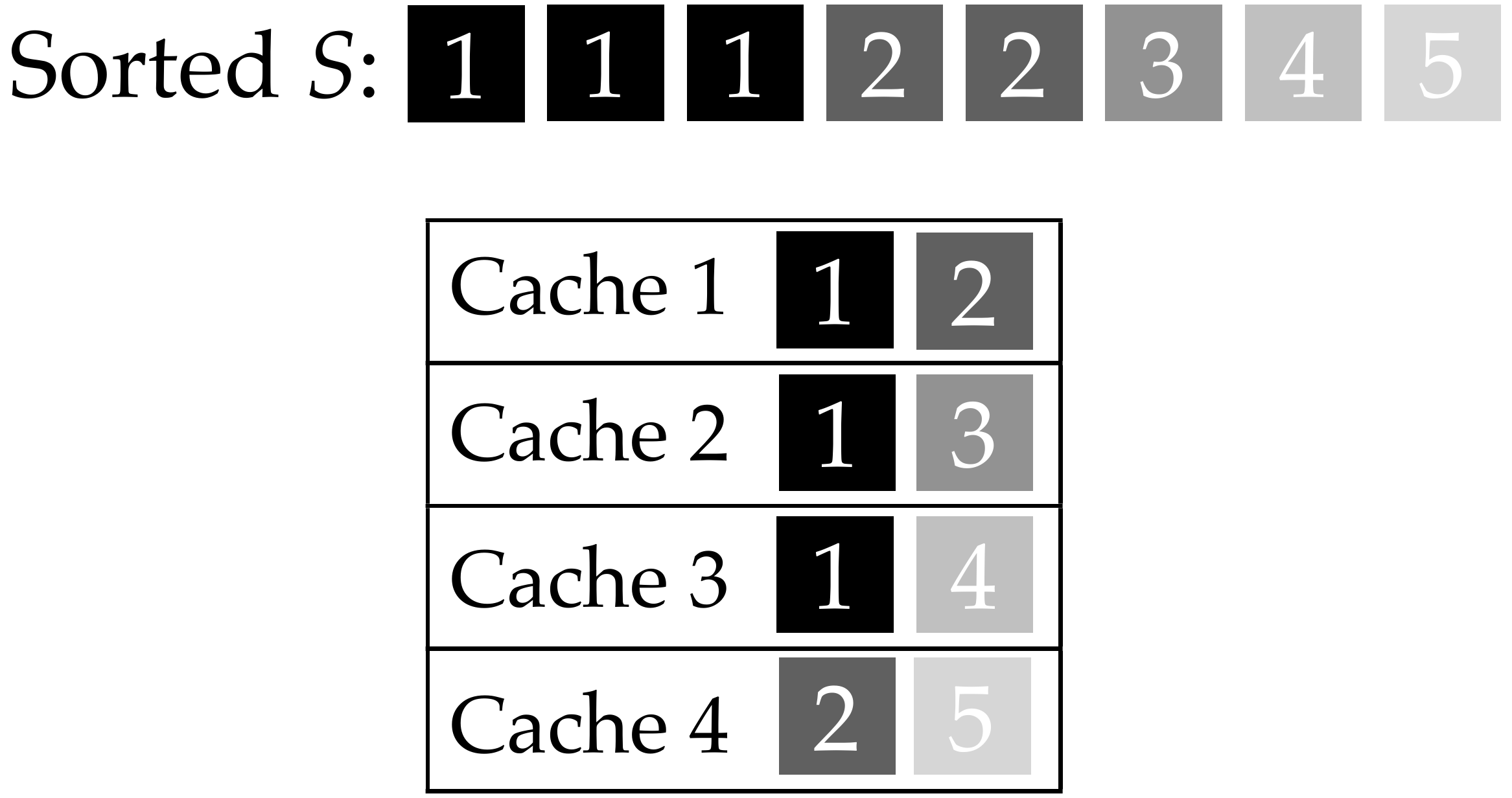}
		\caption{Illustration of Knapsack Storage: Part 2 for a system with four caches ($m=4$).}\label{fig:knapsack_storage_Part 2_example}
	\end{center}
\end{figure}
Figure \ref{fig:knapsack_storage_Part 2_example} illustrates Knapsack Storage: Part 2 for a system consisting of four caches, $x_1 = x_2 = x_3 = x_4 = x_5 = 1$ and 0 otherwise, and $w_1 = 3$, $w_2 = 2$, $w_3 = w_4 = w_5 = 1$.
\begin{remark}
	Knapsack Storage: Part 2 can be implemented in $\OO(n)$ time, where $n$ is the number of contents in the content catalog. 
\end{remark}

\subsubsection{Matching Policy: Match Least Popular}
The next task is to match requests to caches. The key idea of the Match Least Popular policy is to match requests for the less popular contents before matching requests for the more popular contents. Please refer to Figure \ref{fig:match_least_popular} for a formal description of the Match Least Popular policy.
\begin{figure}[h]
	\hrule
	\vspace{0.1in}
	\begin{algorithmic}[1]
		\STATE initialize $i = n$, set of idle caches $= \{1,2,..., m\}$. 
		\IF {the number of requests for Content $i$  is more than the number of idle caches storing Content $i$,}
		\STATE goto Step 8.
		\ELSE 
		\STATE match requests for Content $i$ to idle caches storing Content $i$, chosen uniformly at random. 
		\STATE update the set of idle caches. 
		\ENDIF
		\STATE $i = i-1$, goto Step 2.
	\end{algorithmic}
	\vspace{0.1in}
	\hrule
	\caption{Match Least Popular -- \sl Matches requests to caches.}
	\label{fig:match_least_popular}
\end{figure}
\begin{remark} Match Least Popular can be implemented by going through the set of $m$ requests twice as follows: the first pass is used to tabulate how many times each content is requested, and the second pass is used to match these requests to caches in increasing order of content popularity. Therefore, the Match Least Popular policy can be implemented in $\OO(m)$ time where $m$ is the number of caches as well as the number of requests arriving in each time-slot.
\end{remark}

The next theorem evaluates the performance of the Knapsack Store + Match Least Popular (KS+MLP) policy for the case where content popularity follows the Zipf distribution (Assumption \ref{assumption:Zipf_popularity}). We focus on this case because empirical studies of many VoD services have shown that the content popularity profile matches well with such distributions, see e.g., \cite{liu2013measurement,liu2012server,BC99,YZ06,IRF04,VA02,fricker2012impact}. 

\begin{theorem}
	\label{theorem:performance:our_policy}
	Consider a distributed cache consisting of a central server and $m$ caches that offers a catalog of $n$ contents. Let a batch of $m$ requests arrive in each time-slot and $R_{\text{KS+MLP}}$ be the transmission rate for the Knapsack Store + Match Least Popular policy when content popularity follows the Zipf distribution (Assumption \ref{assumption:Zipf_popularity}) with Zipf parameter $\beta>1$. Let the number of requests arriving in each time-slot be equal to the number of caches. 
	Then, we have that, for $n$ large enough, 
	\begin{eqnarray*}
		R_{\text{KS+MLP}} &\leq& \sum_{i \notin R} 1-\bigg(1-\frac{p_1}{i^{\beta}}\bigg)^m \text{ w.h.p., and,}  \\
		\EE[R_{\text{KS+MLP}}] &\leq& \sum_{i \notin R} 1-\bigg(1-\frac{p_1}{i^{\beta}}\bigg)^m + \OO(n^2 e^{-(\log n)^{2}}),
	\end{eqnarray*}
	where $p_1 = \big(\sum_{i=1}^n i^{-\beta}\big)^{-1}$, $\mathcal{R} = \{i: x_i = 1\}$, such that $x_i$ is the solution of the fraction knapsack problem solved in Knapsack Storage: Part 1, and w.h.p. means with probability $\geq 1-\OO(n e^{-(\log n)^2})$.
\end{theorem}

Table \ref{table:KS+MLP_Zipf_ratio} shows the results for the case where $b_i = b$ $\forall i$, and content popularity follows the Zipf distribution with Zipf Parameter $\beta$, such that $1 < \beta <2$. Typical values of $\beta$ for most Video on Demand services lie between 0.6 and 2. For $\beta<1$, the Zipf distribution is not well defined in the limit as $n \rightarrow \infty$, and therefore, for the asymptotic analysis, we focus our attention on the case where $1 < \beta <2$. In Section \ref{section:simulations}, we evaluate the performance of the Knapsack Storage + Match Least Popular for the case where $0.6<\beta<1$ for finite values of $n$ via simulations. 

\begin{table}
	\begin{center}
		\begin{tabular}{| l| c|}
			\hline
			& $\EE[R_{\text{KS+MLP}}]$ \\ 
			\hline
			Case 1: $ c-\tilde{k} = \Omega(1)$ & $\OO(n^{2-\beta}) \text{ w.h.p.}$ \\
			Case 2: $ c-\tilde{k} = \Theta(n^{-\epsilon})$, $0 <\epsilon \leq \tilde{\epsilon}$ & $\OO(n^{2-\beta - \epsilon}) \text{ w.h.p.}$ \\
			Case 3: $ |\tilde{k}-c| = \OO(n^{-\tilde{\epsilon}})$, & $\OO\big(n^{\frac{2-\beta}{\beta}}\big) \text{ w.h.p.}$ \\
			Case 4: $ \tilde{k}-c = \Theta(n^{-\epsilon})$, $0 <\epsilon \leq \tilde{\epsilon}$ & $\OO(n^{\frac{\epsilon}{\beta-1}}) \text{ w.h.p.}$ \\
			Case 5: $\tilde{k}-c \geq 1 - \oo(n^{1/\beta})$ & $\Theta(1) \text{ w.h.p.}$ \\ 
			\hline
		\end{tabular}
	\end{center}
	\caption{Transmission rate ftom the central server for the Knapsack Storage + Match Least Popular (KS+MLP) policy for the case where content popularity follows the Zipf distribution with Zipf Parameter $\beta$, such that $1 < \beta <2$. Here, $\tilde{\epsilon} = (2-\beta)(\beta-1)/\beta$, each cache can store $\tilde{k}$ contents, and $c = \frac{\text{Number of contents}}{\text{Number of caches}}$. }\label{table:KS+MLP_Zipf_ratio}
	
\end{table}

From Tables \ref{table:Zipf_ratio} and \ref{table:KS+MLP_Zipf_ratio}, we conclude that if content popularity follows the Zipf's distribution with $1 < \beta < 2$, the Knapsack Storage + Match Least Popular policy is order-optimal in the class of policies which do not use coded caching. Using Theorems \ref{theorem:converse} and \ref{theorem:performance:our_policy}, it can be shown that this holds even for Zipf parameter $\beta \geq 2$. This is one of the key results of this paper.

All our results can easily be extended to the case where content popularity follows the Zipf-Mandelbrot law instead of the Zipf's law. We skip the details due to lack of space. 

\subsection{Alternative Distributed Caching Settings}	
We now compare various distributed caching settings studied thus far. In all these setting, the goal is to minimize the rate of transmission from the central server to the users. For each setting, the questions we are trying to answer are:
\begin{enumerate}
	\item[Q1:] What is the optimal caching scheme?
	\item[Q2:] Can coded caching significantly improve performance?
\end{enumerate}
We refer to our setting as Setting A. As discussed in the introduction, there are two other caching settings (Settings B and C) studied so far. We include a possible fourth setting (Setting D) for the sake of completeness. 
\begin{itemize}
	\item[-] \textbf{Setting A:} Flexible matching between users and caches, the central server serves requests which cannot be served by the caches via the root node (our setting).
	\item[-] \textbf{Setting B:} Fixed matching between users and caches, the central server broadcasts information to all the users \cite{maddah2014fundamental, maddah2013decentralized, pedarsani2014online, niesen2014coded, hachem2014multi, zhang2015coded }.
	\item[-] \textbf{Setting C:} Flexible matching between users and caches, independent unicast communication between each user and the central server \cite{LLM12, LLM13, SGSS14, SGSS14_2}.
	\item[-] \textbf{Setting D:} Fixed matching between users and caches, independent unicast communication between each user and the central server.
\end{itemize}

\subsubsection{Setting A- Our Setting}

Theorem \ref{theorem:converse} provides a lower bound on the transmission rate of all policies which do not use coded caching. If content popularity follows the Zipf distribution, our next result provides a stronger information-theoretic converse which lower bounds the performance of all caching policies, including those which employ coded caching. 

\begin{theorem}
	\label{theorem:converse_information_theoretic}
	Let content popularity follow the Zipf distribution with Zipf parameter $\beta > 1$, and $\EE[R^*_{\text{Zipf}}]$ denote the minimum expected transmission rate required to serve all requests arriving in a time-slot. Then, we have that
	$$
	\EE[R^*_{\text{Zipf}}] \ge \Omega\left(\frac{(n - mk)m}{n^{\beta}}\right). 
	$$
\end{theorem}

We now use Theorems \ref{theorem:performance:our_policy} and \ref{theorem:converse_information_theoretic} to compare the performance of our policy with the performance of the optimal policy. 
\begin{corollary}
	Let content popularity follow the Zipf distribution with Zipf parameter $\beta>1$. Let $n=cm$, $\tilde{m} = m$ and 
	each cache have the storage capacity to store exactly $\tilde{k}$ complete equal sized contents. From Theorems \ref{theorem:performance:our_policy} and \ref{theorem:converse_information_theoretic} we have that, $\exists$ constants $c_1$ and $c_2$ such that: \\
	\noindent Case 1: If $\tilde{k} \leq \lceil c \rceil -1$, $\EE[R_{\text{KS+MLP}}]  \leq c_1 \EE[R^*_{\text{Zipf}}].$ \\
	\noindent Case 2: If $\tilde{k} > \lceil c \rceil$, $\EE[R_{\text{KS+MLP}}]  \leq c_2.$
\end{corollary}

\begin{remark}
	We thus conclude that for Setting A, if content popularity follows the Zipf's law, the Knapsack Storage + Match Least Popular policy, which does not use coded caching is optimal up to a constant multiplicative factor and a constant additive term for $\tilde{k} \leq \lceil c \rceil -1$ and $\tilde{k} > \lceil c \rceil$. This is one of the key results of this paper. If coding can reduce the required transmission rate in the case where $\tilde{k} = \lceil c \rceil$ remains an open question. 
\end{remark}

\subsubsection{Setting B- Fixed Matching with Broadcast}
The optimal expected broadcast rate (up to a constant multiplicative and additive gap) for this setup, with an arbitrary popularity distribution, is given by \cite{zhang2015coded}[Theorem 1, Theorem 2]. The proposed scheme uses coded caching and achieves an expected broadcast rate of 
$$
\EE[R(\tilde{k})] \le \left[ \frac{n_3}{\tilde{k}} - 1\right]^+  + \min \left\{ \sum_{i > n_3} mp_i, \ \frac{n - n_3}{[\tilde{k} - n_3]^+} - 1 \right\},
$$
where each cache can store $\tilde{k}$ contents and $n_3$ is an integer that satisfies $m\tilde{k}p_{n_3} \ge 1$ and $m\tilde{k}p_{n_3+1} < 1$. Here, $n_3$ is the number of contents for which are stored in the caches \cite{maddah2014fundamental}. Consider the Zipf distribution with Zipf parameter $\beta > 1$. Then $n_3 = \Theta \left(\min\{(\tilde{k}m)^{1/\beta}, n\}\right)$, and therefore the achievable expected rate is given by
\begin{align}
	\nonumber
	\EE[R(\tilde{k})] &\le \OO\left( \frac{ (\tilde{k}m)^{1/\beta} }{\tilde{k}} + m \sum_{i > n_3} i^{-\beta} \right)\\
	&\le \OO\left( \frac{ m^{1/\beta} }{\tilde{k}^{1 - 1/\beta}} + m (\tilde{k}m)^{-(\beta - 1)/ \beta} \right) = \OO\left( \frac{ m^{1/\beta} }{\tilde{k}^{1 - 1/\beta}} \right).
	\label{Eqn:Coded}
\end{align}

The conventional uncoded caching and delivery scheme would be to store the most popular $\tilde{k}$ contents in each cache and simply broadcast those requested contents which have not been stored. The expected broadcast rate for this scheme is 
$$
\EE[R_{\text{NC}}(\tilde{k})] = \sum_{i > \tilde{k}} mp_i . 
$$
Therefore, for the Zipf distribution with $\beta > 1$, 
\begin{equation}
	\label{Eqn:Uncoded}
	\EE[R_{\text{NC}}(\tilde{k})]  \ge \Omega\left( m \cdot \tilde{k}^{-(\beta - 1)} \right) .
\end{equation}
Consider $\tilde{k} = m^{1/\beta}$. Comparing the coded caching and conventional caching rates using \eqref{Eqn:Coded} and \eqref{Eqn:Uncoded}, we have 
\begin{eqnarray*}
	\frac{\EE[ R_{\text{NC}}(m^{1/\beta})] }{ \EE[R(m^{1/\beta})] } &\ge&  \Omega\left( m \cdot m^{-\frac{1}{\beta}(\beta - 1)} \cdot \frac{m^{\frac{1}{\beta}(1 - \frac{1}{\beta})}}{ m^{1/\beta} } \right)\\
	&=&  \Omega\left(  m^{\frac{1}{\beta}(1 - \frac{1}{\beta})} \right), 
\end{eqnarray*}
which can grow arbitrarily large as $m \rightarrow \infty$. Therefore, for this setting, optimal performance cannot be achieved without using coded caching, and the optimal policy stores the more popular contents. 

This result can be extended to the case where content popularity follows the Zipf-Mandelbrot law. We skip the details due to lack of space.
\subsubsection{Setting C- Flexible Matching with Unicast}

We first focus on the case where each cache can store exactly one content, i.e., $\tilde{k} = 1$. The next theorem provides a lower bound on the transmission rate for all policies which do not use coded caching. 
\begin{theorem}
	\label{theorem:setting_C_converse}
	For Setting C, let $\mathcal{C}^{*}(1)$ be the minimum transmission from the central server needed to serve all requests which arrive in a time-slot if each cache has sufficient storage capacity to store exactly one complete content. If each request is generated by an i.i.d. process such that the probability of a request for Content $i$ is $p_i$, then we have that,
	\begin{eqnarray*}
		\EE[{\mathcal{C}^{*}(1)}] &=& \tilde{m}- O_c, \\
		\text{where } O_c &=& \max \sum_{i=1}^n \sum_{j=1}^{\tilde{m}} y_{i,j} v_{i,j} \\
		&& s.t. \sum_{i=1}^n \sum_{j=1}^{\tilde{m}} y_{i,j} = m, \\
		&& y_{i,j} = \{0,1 \},\text{ } \forall i,j,
	\end{eqnarray*}
\end{theorem}
where $v_{i,j} = \sum_{k=j}^{\tilde{m}} {\tilde{m} \choose k} p_i^k (1-p_i)^{\tilde{m}-k}$ is the probability that Content $i$ is requested at least $j$ times in a batch of requests. 

Consider the following storage policy: store at most one content per cache with Content $i$ stored on $c_i$ caches, such that, $c_i = \sum_{i=1}^{\tilde{m}} y_{i,j}, $
where the $y_{i,j}s$ are the solution to the following optimization problem:
\begin{eqnarray*}
	\max &&\sum_{i=1}^n \sum_{j=1}^{\tilde{m}} y_{i,j} v_{i,j} \\
	&& s.t. \sum_{i=1}^n \sum_{j=1}^{\tilde{m}} y_{i,j} = m, \\
	&& y_{i,j} = \{0,1 \},\text{ } \forall i,j.
\end{eqnarray*}
Note that the expected transmission from the central server for this policy matches the minimum expected transmission rate obtained in Theorem \ref{theorem:setting_C_converse}. It follows that this policy is optimal for Setting C if each cache can store exactly one content. Since $v_{i_1,j} \geq v_{i_2,j}$ if $p_1 \geq p_2$, this policy stores the more popular contents on a larger number of caches.  

\begin{corollary}
	\label{corollary:proportional_storage}
	For Setting C, if content popularity follows the Zipf popularity (Assumption 1), $\tilde{m} = m$, and each cache can store exactly one complete content, under the optimal policy, the expected number of unicast transmission required from the central server in a time-slot is $\OO(m^{1/\beta} \log m)$.
\end{corollary}

From Theorem \ref{theorem:performance:our_policy}, we know that if content popularity follows the Zipf's law, the Knapsack Storage + Match Least Popular policy is optimal up to a constant additive term for Setting A. Next, we evaluate the performance of the Knapsack Storage + Match Least Popular policy for Setting C.

\begin{corollary}
	\label{corollary:knapsack_setting_C}
	Let content popularity follows the Zipf popularity (Assumption 1) with Zipf parameter $\beta>1$, $\tilde{m} = m$, and each cache has sufficient storage capacity to store $\tilde{k} \geq 1$ complete contents. For Setting C, under the Knapsack Storage + Match Least Popular policy, for $n$ large enough, 
	\begin{enumerate}
		\item If $\tilde{k} = 1$, the expected number of unicast transmissions required from the central server in a time-slot is $\Theta(m)$.
		\item If $\tilde{k} \geq \lceil c \rceil+2$, the expected number of unicast transmission required from the central server in a time-slot is $\oo(1)$.
	\end{enumerate}
\end{corollary}

\begin{remark}
	From Corollaries \ref{corollary:proportional_storage} and \ref{corollary:knapsack_setting_C}, we conclude that for Setting C, if content popularity follows the Zipf distribution with Zipf parameter $\beta>1$: 
	\begin{enumerate}
		\item[--] if $\tilde{k} = 1$, the Knapsack Storage + Match Least Popular policy is order wise suboptimal. The key thing to note is that for $\tilde{k} = 1$, the Knapsack Storage policy does not cache the most popular content and is outperformed a policy which caches more copies of the more popular contents. 
		\item[--] if $\tilde{k} \geq \lceil c \rceil+2$, the Knapsack Storage + Match Least Popular policy is optimal up to a constant additive term. More specifically, by Theorem \ref{theorem:performance:our_policy}, we have that, with high probability, all requests are served by the local caches by the Knapsack Storage + Match Least Popular policy. Therefore, for the Knapsack Storage + Match Least Popular policy, constant storage per cache (i.e., $\tilde{k} = \lceil c \rceil+2$) is sufficient to ensure $\oo(1)$ transmission rate from the central server. We thus conclude that coded caching is not necessary for optimal performance in this setting. 
	\end{enumerate}
\end{remark}

%

\subsubsection{Setting D- Fixed Matching with Unicast}

In this setting, since each request can only be served by a pre-determined cache and the central server communicates with each user separately, the distributed caching problem decouples into $m$ independent single cache problems where the goal is to minimize the probability that the requested content is not stored in the cache. It is straightforward to see that in this setting if a cache can store $\tilde{k}$ complete contents, the optimal caching policy which does not use coding is to store the $\tilde{k}$ most popular contents.

\section{Simulations}
\label{section:simulations}
In this section, we compare the performance of the Knapsack Storage + Match Least Popular (KS+MLP) policy with the lower bound on the performance of all storage policies which do not use coded caching.

We simulate a distributed cache system for arrival processes which satisfy Assumption \ref{assumption:Zipf_popularity} to understand how the performance of the KS+MLP policy depends on various parameters like number of contents $(n)$, number of caches ($m$), storage capacity per cache $(\tilde{k})$, and Zipf parameter $(\beta)$. We focus on the case where the number of requests per time-slot is equal to the number of caches. For each set of system parameters, we report the mean transmission rate averaged over 10000 iterations.

\begin{figure}[h]
	\begin{center}
		\includegraphics[scale=0.27]{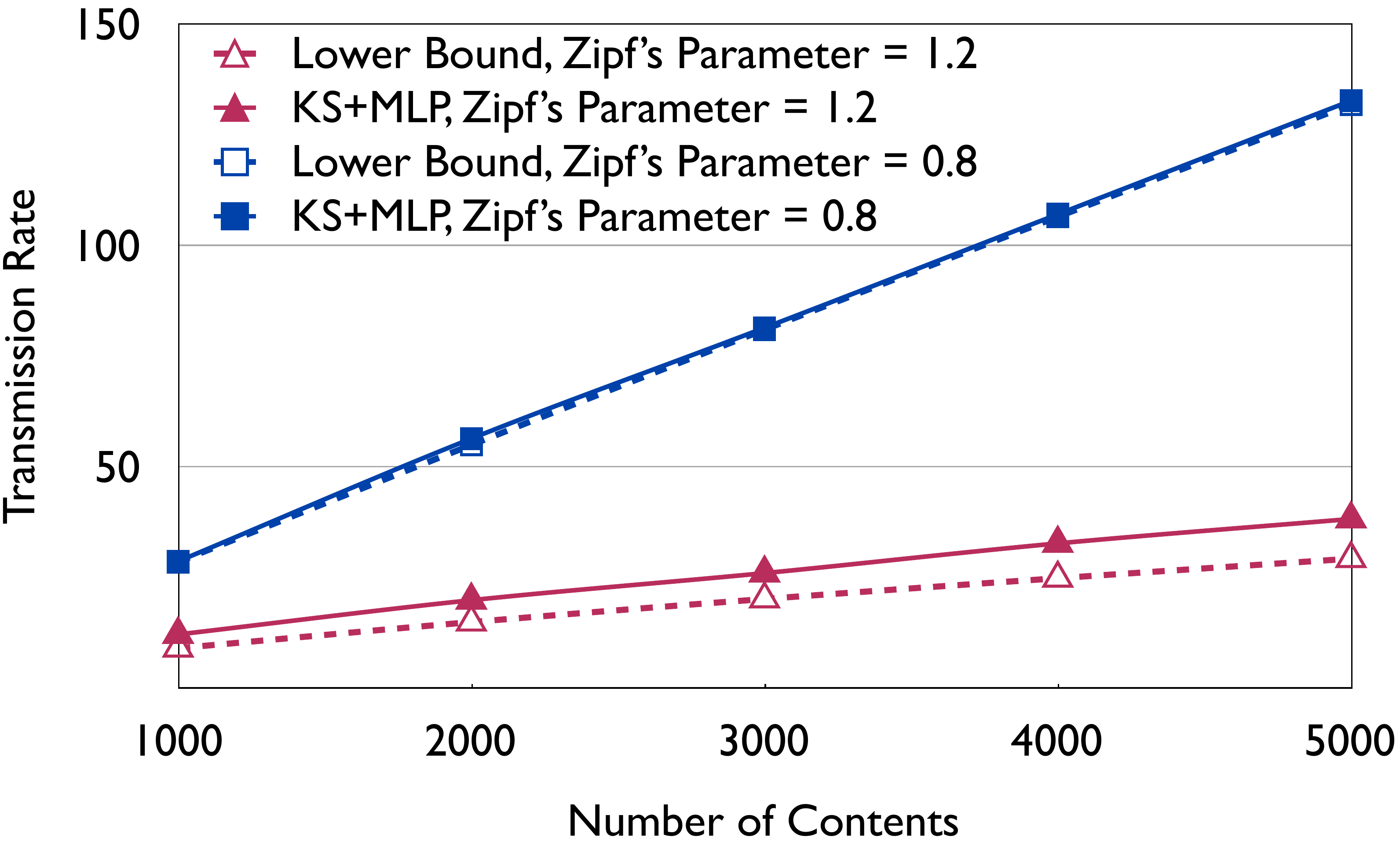}
		\caption{\sl Plot of the mean transmission rate for the KS+MLP policy and the lower bound on the expected transmission rate as a function of the number of contents ($n$), for a system where the number of caches ($m$) is one fifth of the number of contents ($n = 5m$), and each cache can store three contents ($\tilde{k}=3$). \label{fig:vary_size_1}}
	\end{center}
\end{figure}

In Figure \ref{fig:vary_size_1}, we plot the mean transmission rate for the KS+MLP policy and the lower bound on the expected transmission rate as a function of the number of contents ($n$), for a system where the number of caches ($m$) is one fifth of the number of contents ($n = 5m$), and each cache can store three contents ($\tilde{k}=3$). In this regime, our theoretical results suggest that the mean transmission rate for the KS+MLP policy is $\OO(n^{2-\beta})$ and the lower bound on the expected transmission rate to be $\Omega(n^{2-\beta})$. We see that the mean transmission rate for the KS+MLP policy is very close to the lower bound and both are decreasing functions of $\beta$. 

\begin{figure}[h]
	\begin{center}
		\includegraphics[scale=0.27]{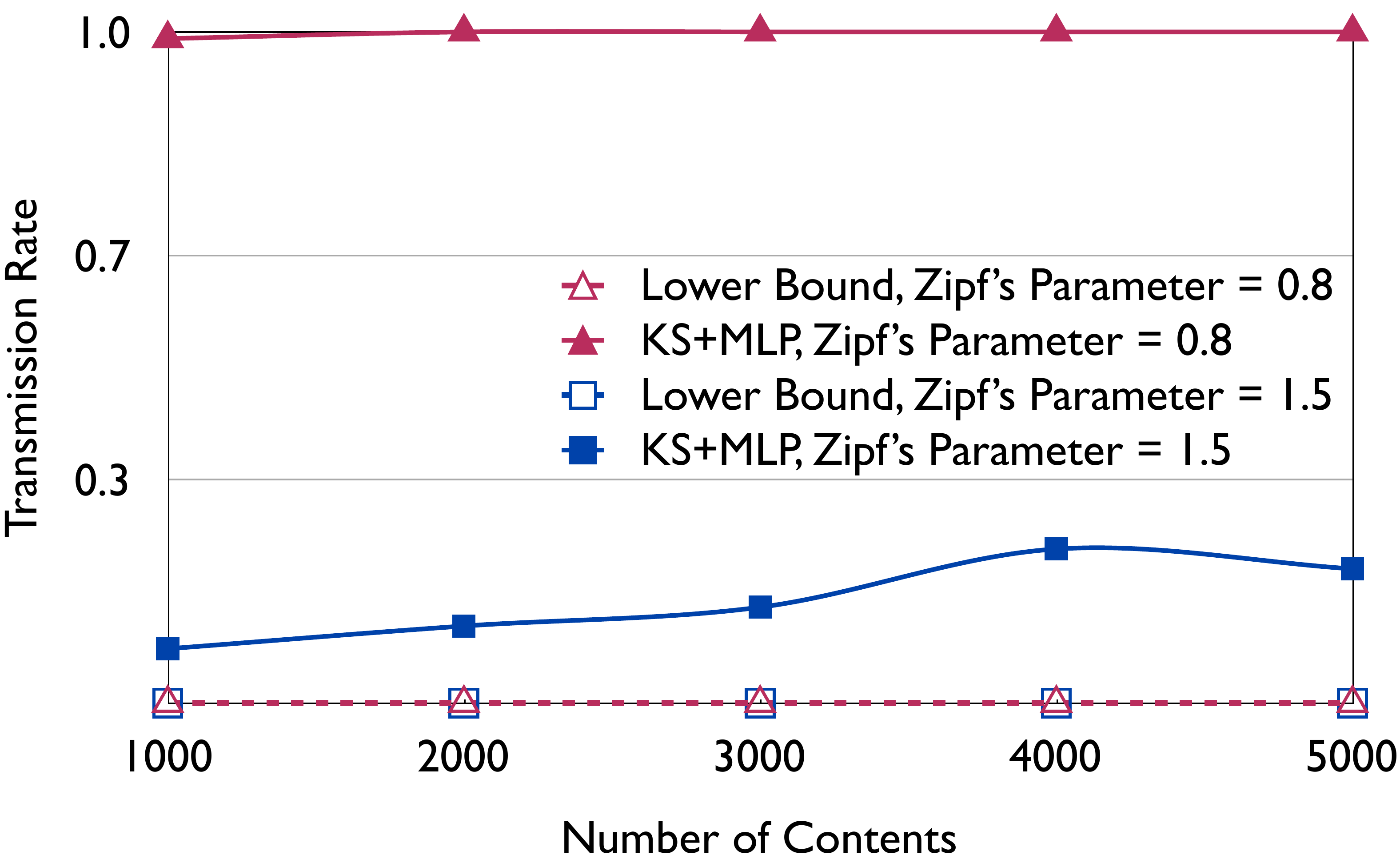}
		\caption{\sl Plot of the mean transmission rate for the KS+MLP policy and the lower bound on the expected transmission rate as a function of the number of contents ($n$), for a system where the number of caches, $m = n/15$, and each cache can store sixteen contents ($\tilde{k}=16$). \label{fig:vary_size_2}}
	\end{center}
\end{figure}

In Figure \ref{fig:vary_size_2}, we plot the mean transmission rate for the KS+MLP policy and the lower bound on the expected transmission rate as a function of the number of contents ($n$), for a system where $n = 15 m$ and each cache can store sixteen contents ($\tilde{k}=16$). In this regime, our theoretical results suggest that the mean transmission rate for the KS+MLP policy is upper bounded by one, with high probability and the lower bound on the expected transmission rate is zero. 

\begin{figure}[h]
	\begin{center}
		\includegraphics[scale=0.27]{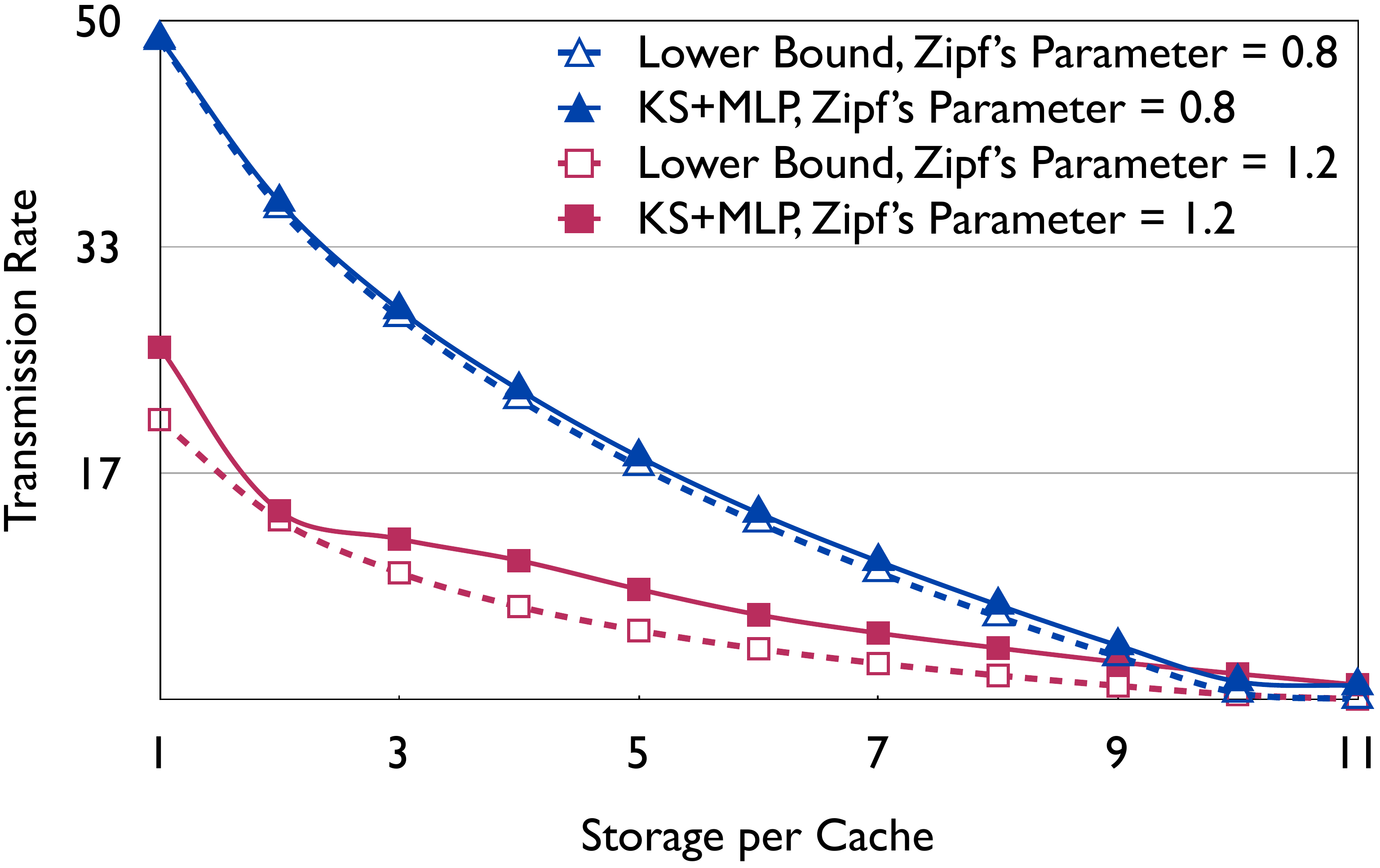}
		\caption{\sl Plot of the mean transmission rate for the KS+MLP policy and the lower bound on the expected transmission rate as a function of storage capacity per cache ($\tilde{k}$) for a system with 1000 contents ($n=1000$) and 100 caches ($m=100$). \label{fig:vary_storage}}
	\end{center}
\end{figure}

In Figure \ref{fig:vary_storage}, we plot the mean transmission rate for the KS+MLP policy and the lower bound on the expected transmission rate as a function of the storage per cache ($\tilde{k}$) for a system with 1000 contents ($n=1000$) and 100 caches ($m=100$). We see that the mean transmission rate for the KS+MLP policy is very close to the lower bound on the expected transmission rate and as expected, both are decreasing function of the storage per cache ($\tilde{k}$).

\begin{figure}[h]
	\begin{center}
		\includegraphics[scale=0.27]{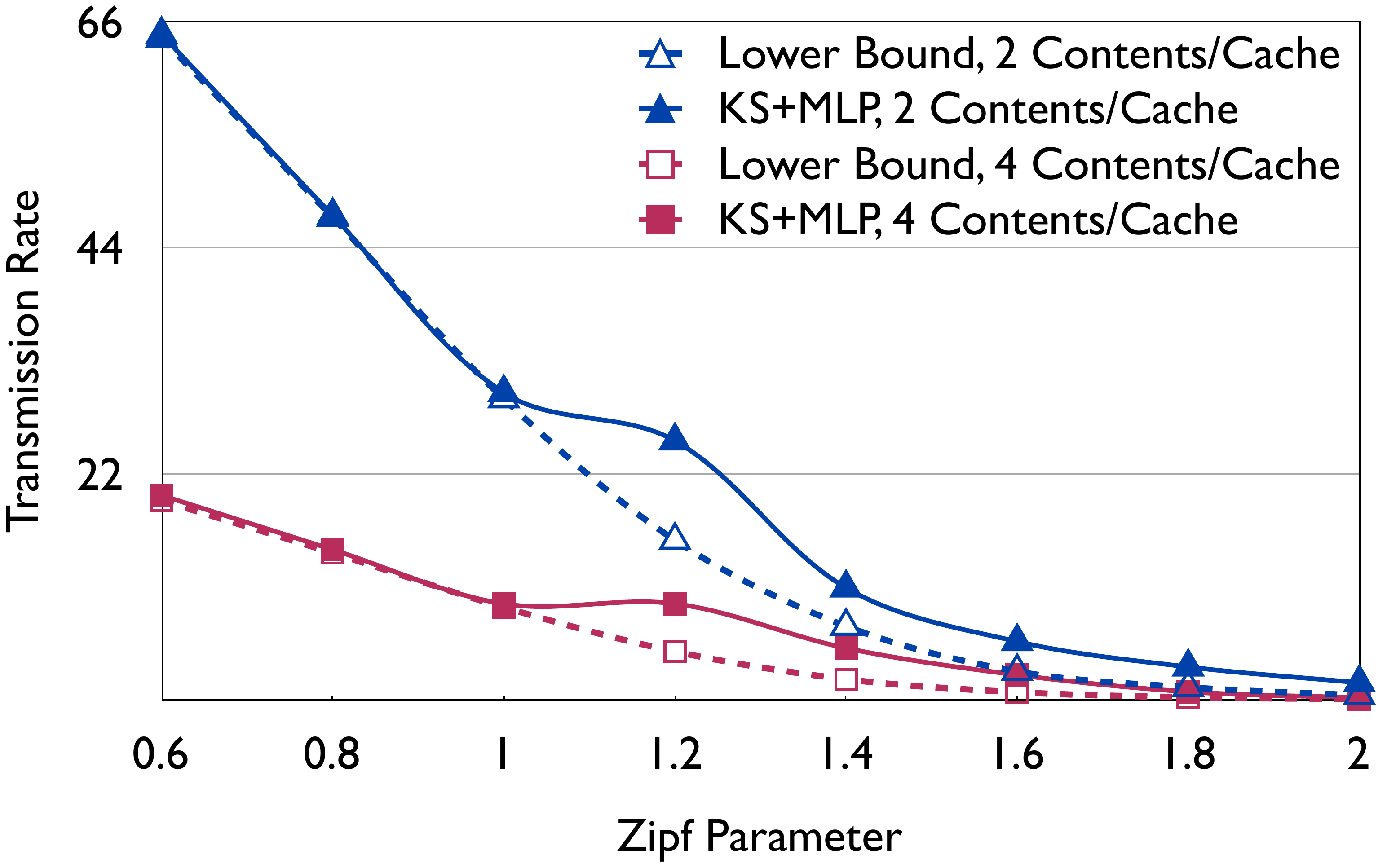}
		\caption{\sl Plot of the mean transmission rate for the KS+MLP policy and the lower bound on the expected transmission rate as a function of Zipf parameter ($\beta$) for a system with 1000 contents ($n=1000$) and 200 caches ($m=200$). \label{fig:vary_beta}}
	\end{center}
\end{figure}

In Figure \ref{fig:vary_beta}, we plot the mean transmission rate for the KS+MLP policy and the lower bound on the expected transmission rate as a function of the Zipf parameter $\beta$. Typical values of $\beta$ for video of demand systems lie between 0.6 and 2 \cite{liu2013measurement,liu2012server,BC99,YZ06,IRF04,VA02,fricker2012impact}. We simulate a system with 1000 contents ($n=1000$) and 200 caches ($m=200$) for two different values of storage per cache. We see that the mean transmission rate for the KS+MLP policy is very close to the lower bound on the expected transmission rate and as expected, both the mean transmission rate for the KS+MLP policy and the lower bound on the expected transmission rate are decreasing functions of $\beta$. 

\section{Proofs}
\label{section:proofs}

In this section, we prove some of the results discussed in Section \ref{section:main_results}. 
We use the following result multiple times in this section.

\begin{lemma}
	\label{lemma:chernoff}
	For a Binomial random variable $X$ with mean $\mu$, by the Chernoff bound, for $\delta$ such that $0 < \delta < 1$,
	\begin{eqnarray*}
		&& \PP(X \geq (1+\delta) \mu) \leq e^{-\delta^2 \mu/3}, \\
		&& \PP(X \leq (1-\delta) \mu) \leq e^{-\delta^2 \mu/2}.
	\end{eqnarray*}
\end{lemma}

\subsection{Proof of Theorem \ref{theorem:converse}}


\begin{IEEEproof}[Proof of Theorem \ref{theorem:converse}]
	Instead of lower bounding the expected transmission rate for the system described in Section \ref{section:system_model}, we lower bound the expected transmission rate for an alternative system which is less restrictive and, therefore, more powerful than the system described in Section \ref{section:system_model}. In the system described in Section \ref{section:system_model}, each cache can serve at most one request in a time-slot. In the alternative system, we allow each cache to serve multiple requests in each time-slot as long as it serves at most one request for each content stored in the cache. 
	
	Let $\EE[R^*_{\text{NC}}]$ be the expected transmission rate for the system described in Section \ref{section:system_model}, and $\EE[\tilde{R}^*_{\text{NC}}]$ be the expected transmission rate for the alternative system.
	Since the alternative system is less restrictive than the original system, it follows that,
	$$\EE[R^*_{\text{NC}}] \geq \EE[\tilde{R}^*_{\text{NC}}].$$
	
	In the alternative system, for Content $i$ such that $\tilde{m}p_i \geq 1$, if the caching policy decides to store bit $v$ of Content $i$ on less than $\tilde{m}p_i$ caches, the caches cannot serve $\tilde{m}p_i$ or more requests for Content $i$ in any given time-slot. Since the number of requests for Content $i$ is at least $\tilde{m}p_i$ with a constant probability, storing bit $v$ of Content $i$ such that $\tilde{m}p_i \geq 1$ on less than $\tilde{m}p_i$ caches adds a constant to the expected transmission rate and not storing bit $v$ of Content $i$ on the caches adds one unit to the expected transmission rate. Since we are interested in the order of the expected transmission rate as $n, m, \tilde{m} \rightarrow \infty$, if $\tilde{m} p_i \geq 1$, storing bit $v$ of Content $i$ on less than $\tilde{m} p_i$ caches is equivalent to not storing bit $v$ of Content $i$ at all. Therefore, to make the most use of the available cache memory, we restrict ourselves to the case where if the caching policy decides to cache bit $v$ of Content $i$, it is stored on at least $\max\{\tilde{m} p_i,1\}$ caches. 
	
	If the caching policy decides not to cache bit $v$ of Content $i$, this bit will have to be transmitted if Content $i$ is requested at least once in the batch of $\tilde{m}$ requests. The probability that Content $i$ is requested at least once in the batch of $\tilde{m}$ requests is $1-(1-p_i)^{\tilde{m}}$. Let $x_{i,v}$ be an indicator variable, where $x_{i,v}=1$ implies that bit $v$ of Content $i$ is cached and $x_{i,v}=0$ otherwise. Since the system has $m$ caches where each cache can store $k$ bits of content, the expected transmission rate for the alternative system ($\EE[\tilde{R}^*_{\text{NC}}]$) is lower bounded by the solution to the following optimization problem:
	\begin{eqnarray*}
		\min && \displaystyle \sum_{i=1}^n \displaystyle \sum_{u=1}^{b_i} (1-x_{i,u}) (1-(1-p_i)^{\tilde{m}}) \\
		\text{s.t. } && \displaystyle \sum_{i=1}^n \displaystyle \sum_{u=1}^{b_i} x_{i,v} \max\{\tilde{m}p_i,1\} \leq mk, \\
		&& x_{i,u} = \{0,1\}, \text{ } \forall i,u.
	\end{eqnarray*}
	Let $\text{O}_1^* =$
	\begin{eqnarray*}
		\max && \displaystyle \sum_{i=1}^n \displaystyle \sum_{u=1}^{b_i} x_{i,u} (1-(1-p_i)^{\tilde{m}}) \\
		\text{s.t. } && \displaystyle \sum_{i=1}^n \displaystyle \sum_{u=1}^{b_i} x_{i,u} \max\{\tilde{m}p_i,1\} \leq mk, \\
		&& x_{i,u} = \{0,1\}, \text{ } \forall i,u.
	\end{eqnarray*}
	To lower bound $\EE[\tilde{R}^*_{\text{NC}}]$, we need to upper bound $\text{O}_1^*$. Instead of upper bounding $\text{O}_1^*$, we upper bound $\text{O}^*$, where $\text{O}^* =$
	\begin{eqnarray*}
		\max && \displaystyle \sum_{i=1}^n \displaystyle \sum_{u=1}^{b_i} x_{i,u} (1-(1-p_i)^{\tilde{m}}) \\
		\text{s.t. } && \displaystyle \sum_{i=1}^n \displaystyle \sum_{u=1}^{b_i} x_{i,u} \max\{\tilde{m}p_i,1\} \leq mk, \\
		&& 0 \leq x_{i,u} \leq 1, \text{ } \forall i,u.
	\end{eqnarray*}
	By definition, $\text{O}_1^* \leq \text{O}^*$, therefore, 
	\begin{eqnarray*}
		\EE[R^*_{\text{NC}}] \geq \EE[\tilde{R}^*_{\text{NC}}] \geq \displaystyle \sum_{i=1}^n b_i (1-(1-p_i)^{\tilde{m}}) - \text{O}^*.
	\end{eqnarray*}
\end{IEEEproof}

\subsection{Proof of Corollary \ref{corollary:converse_zipf}} 

Let $\tilde{i}$ be the smallest value of $i$ for which $mp_i < 1$. For all $i \geq \tilde{i}$, $\max\{mp_i,1\} = 1$. Therefore, if content popularity follows the Zipf's law, $\text{O}^* = $
\begin{eqnarray*}
	\max && \displaystyle \sum_{i=1}^n \displaystyle \sum_{v=1}^b x_{i,v} (1-(1-p_i)^m) \\
	\text{s.t. } && \displaystyle \sum_{i=1}^{\tilde{i}-1} \displaystyle \sum_{v=1}^b x_{i,v} mp_i + \displaystyle \sum_{i=\tilde{i}}^n \displaystyle \sum_{v=1}^b x_{i,v} \leq mk, \\
	&& 0 \leq x_{i,v} \leq 1, \text{ } \forall i,v.
\end{eqnarray*}
$\text{O}^*$ is the solution to the fractional knapsack problem where the value of bit $u$ of content $i$ is $1-(1-p_i)^m$ and its weight is $mp_i$ if $i \leq \tilde{i}-1$ and 1 if $i \geq \tilde{i}$. Let $r_i$ be the value to weight ratio of a bit of content $i$. We have that, 
\begin{eqnarray*}
	r_i = \frac{v_i}{w_i} = \begin{cases}
		\dfrac{1-(1-p_i)^m}{mp_i}, & \text{ for } i \leq \tilde{i}-1, \\
		1-(1-p_i)^m, & \text{   for } i \geq \tilde{i}.
	\end{cases}
\end{eqnarray*}
Given this,
\begin{eqnarray*}
	\frac{\text{d}r_i}{\text{d}i} >0,  \text{ for } i \leq \tilde{i}-1 \text{, and } \frac{\text{d}r_i}{\text{d}i} <0,  \text{ for } i \geq \tilde{i}.
\end{eqnarray*}

Therefore, $r_i$ increases from $i=1$ to $\tilde{i}-1$ and decreases from $i=\tilde{i}$ to $n$. Therefore, the optimal solution to the fraction knapsack solution has the following structure: \\
$\exists$ $i_{\text{min}}$, $i_{\text{max}}$ with $i_{\text{min}} \leq \tilde{i} \leq i_{\text{max}}$, such that, either,
\begin{eqnarray*}
	\sum_{i=i_{\text{min}}}^{i_{\text{max}} -1} \max\{mp_i,1\} \leq mk, \text{ and}  \sum_{i=i_{\text{min}}}^{i_{\text{max}}} \max\{mp_i,1\} > mk,
\end{eqnarray*}
and the optimal solution is 
\begin{eqnarray*}
	x_{i,u}  = \begin{cases}
		1, & 1 \leq u \leq b \text{, } i_{\text{min}} \leq i \leq i_{\text{max}}-1, \\
		1, & \text{for } i = i_{\text{max}} \text{, } u \leq  mk - \displaystyle\sum_{i=i_{\text{min}}}^{i_{\text{max} -1}} b_i \max\{mp_i,1\}, \\
		0, & \text{otherwise},
	\end{cases}
\end{eqnarray*}
or,
\begin{eqnarray*}
	\sum_{i=i_{\text{min}}+1}^{i_{\text{max}}} \max\{mp_i,1\} \leq mk, \text{ and}  \sum_{i=i_{\text{min}}}^{i_{\text{max}}} \max\{mp_i,1\} > mk,
\end{eqnarray*}
and the optimal solution is 
\begin{eqnarray*}
	x_{i,u}  = \begin{cases}
		1, & 1 \leq u \leq b \text{, } i_{\text{min}}+1 \leq i \leq i_{\text{max}}, \\
		1, & \text{for } i = i_{\text{min}} \text{, } u \leq  mk - \displaystyle\sum_{i=i_{\text{min}}+1}^{i_{\text{max}}} b_i \max\{mp_i,1\}, \\
		0, & \text{otherwise}.
	\end{cases}
\end{eqnarray*}
We use this insight and optimize over the possible values of $i_{\text{min}}$ to compute a lower bound on $\EE[R^*_{\text{NC}}]$ for various values of the number of caches ($m$), number of contents ($n = cm$), storage per cache ($k$) and Zipf parameter ($\beta$). 


We focus on the case where content popularity follows the Zipf distribution with Zipf Parameter $\beta$, with $1 < \beta <2$. Let $i_{min} = m^{\alpha}$. Therefore, 
\begin{eqnarray*}
	&& \sum_{i=m^{\alpha}}^{\tilde{i}} mp_i + \sum_{i=\tilde{i}}^{i_{max}} 1 \leq \frac{mk}{b} \\
	&\Rightarrow& \frac{-1}{\beta-1}\frac{mp_1}{i^{\beta-1}} \bigg|_{\tilde{i}}^{m^{\alpha}}	+ i_{max} - \tilde{i} \leq \frac{mk}{b} \\
	&\Rightarrow& i_{max} \leq \frac{mk}{b} - \frac{p_1m^{1-\alpha(\beta-1)}}{(\beta-1)} + \frac{\beta(p_1m)^{\frac{1}{\beta}} }{(\beta-1)}.
\end{eqnarray*}
The cumulative weight of contents ranked between $i_{max}+1$ to $n$ is $\sum_{i_{max}}^{n} bmp_i$, where,
\begin{eqnarray*}
	\sum_{i_{max}}^{n} bmp_i \geq \frac{1}{\beta-1}\frac{mp_1b}{i^{\beta-1}} \bigg|_{n}^{i_{max}}\geq \frac{1}{\beta-1} \bigg(\frac{mp_1b}{i_{max}^{\beta-1}} -  \frac{mp_1b}{n^{\beta-1}}\bigg).
\end{eqnarray*}
Let $F_1$ be the event that not more than one request arrives for contents ranked between $i_{max}+1$ to $n$. Using the Chernoff bound (Lemma \ref{lemma:chernoff}), $\PP(F_1) \geq \OO(ne^{-(\log n)^2})$. Let $F_2$ be the event that at least one request arrives for contents ranked between one and $i_{min}$. Using the Chernoff bound (Lemma \ref{lemma:chernoff}), $\PP(F_2) \geq \OO(ne^{-(\log n)^2})$. Therefore,
\begin{eqnarray*}
	\EE[R^*_{\text{NC}}|F_1 \cap F_2] &\geq& i_{min}b + \frac{b}{\beta-1} \bigg(\frac{mp_1}{i_{max}^{\beta-1}} -  \frac{mp_1}{n^{\beta-1}}\bigg),\\
	\text{where, }i_{min} &=& m^{\alpha}, \text{ and,}\\
	i_{max} &=& \frac{mk}{b} - \frac{p_1m^{1-\alpha(\beta-1)}}{(\beta-1)} + \frac{\beta(p_1m)^{\frac{1}{\beta}} }{(\beta-1)}.
\end{eqnarray*}
We use the fact that,
\begin{eqnarray*}
	\EE[R^*_{\text{NC}}] \geq \EE[R^*_{\text{NC}}|F_1 \cap F_2] \PP(F_1 \cap F_2),
\end{eqnarray*}
and optimize over $\alpha$ to get the desired result for particular values of $c$, $k$ and $m$. Please refer to Table \ref{table:Zipf_ratio} in Section \ref{section:main_results} for the results.\\

\subsection{Proof of Theorem \ref{theorem:performance:our_policy}}
If content popularity follows the Zipf distribution (Assumption \ref{assumption:Zipf_popularity}) with Zipf parameter $\beta>1$, and $\tilde{m} = m$, the weights of various contents in the Knapsack problem solved by Knapsack Storage: Part 1 are as follows: 
\begin{definition}
	\label{def:knapsack_zipf}
	\begin{eqnarray*}
		n_1 &=& \dfrac{m^{1/\beta}}{p_1(\log m)^{2/\beta}} \text{, } n_2 = \dfrac{m^{1/\beta}(\log m)^{2/\beta}}{p_1}\\
		w_i &=& \begin{cases}
			m & \text{for } i = 1, \\
			\bigg\lceil \bigg(1 + \dfrac{p_1}{2}\bigg) mp_i \bigg\rceil & \text{for } 2 \leq i \leq n_1 \\
			\big\lceil 4 p_1(\log n)^2 \big\rceil & \text{for } n_1+1 \leq i \leq n_2 \\
			1 & \text{for } n_2+1 \leq i \leq n 
		\end{cases} \\
	\end{eqnarray*}
\end{definition}

We use the following two lemmas in the proof of Theorem \ref{theorem:performance:our_policy}. The proofs are discussed in the Appendix. 
\begin{lemma}
	\label{lemma:Zipf_d}
	Let content popularity follow the Zipf distribution with Zipf parameter $\beta>1$. In a given time-slot, let $d_i$ be the number of requests for Content $i$. Let $E_1$ be the event that:
	\begin{enumerate}
		\item[(a)] $d_i \leq 2p_1 (\log n)^2$ for $n_1 < i \leq n_2$,
		\item[(b)] $d_i \leq \bigg(1+\dfrac{p_1}{4}\bigg) mp_i$ for $1 \leq i \leq n_1$,
	\end{enumerate}
	where $n_1$, $n_2$ and $w_i$ are as defined in Definition \ref{def:knapsack_zipf}.
	Then we have that, 
	$$\PP(E_1) = 1-\OO(n e^{-(\log m)^{2}}).$$
\end{lemma}

\begin{lemma}
	\label{lemma:serve_all_requests}
	Let $\mathcal{R} = \{i: x_i = 1\}$, where $x_i$ is the solution of the fraction knapsack problem solved in Knapsack Storage: Part 1. Let $E_2$ be the event that the Match Least Popular policy matches all requests for all contents in $R$ to caches. Then we have that, 
	$$\PP(E_2) = 1-\OO(m e^{-(\log m)^{2}}).$$  
\end{lemma}

\begin{IEEEproof}[Proof of Theorem \ref{theorem:performance:our_policy}]
	From Lemma \ref{lemma:serve_all_requests}, we know that, for $n$ large enough, with probability $\geq 1-\OO\big(me^{-(\log m)^2}\big)$, all requests for the contents caches by the KS+MLP policy are matched to caches. Let $\tilde{n}$ be the number of contents not in $\mathcal{R}$ (i.e., not cached by the KS+MLP policy) that are requested at least once in a given time-slot. Therefore, 
	\begin{eqnarray*}
		\EE[R_{\text{KS+MLP}}] &\leq& \EE[\tilde{n}] P(E_2) + m (1-P(E_2)) \\
		&\leq& \EE[\tilde{n}] + \OO(m^2 e^{-(\log m)^{2}}). 
	\end{eqnarray*}
	
	The results follow by solving the fractional knapsack problem defined in Figure \ref{fig:knapsack_storage_Part 1} as a function of $n$, $m$, $\beta$ and $\tilde{k}$ to determine the set $\mathcal{R}$. For a given set $\mathcal{R}$, 
	$$\EE[\tilde{n}] = \sum_{i \notin \mathcal{R}} 1-(1-p_1)^m.$$
	We omit the details due to lack of space. 
\end{IEEEproof}

\subsection{Proof of Theorem \ref{theorem:converse_information_theoretic}}

\begin{IEEEproof}[Proof of Theorem \ref{theorem:converse_information_theoretic}]
	In our setup, say System I, there are $m$ users and $m$ caches, each with a storage capacity of $k$ bits. In each time-slot, each user requests for a content according to the Zipf distribution with parameter $\beta$. The least popular content (Content $n$) has popularity $p_n = \OO(n^{-\beta})$. We are interested in establishing a lower bound on the expected rate for this system. 
	
	Consider an alternative setup, say System II, where we introduce a dummy empty content $W_{0}$. Each user requests Content $i$ for $1 \leq i \leq n$ with equal probability $p = \OO(n^{-\beta})$ and the empty content $W_{0}$ with probability $1 - \OO(n^{1-\beta})$. Using the fact that each non-empty content is requested with higher probability in System I as opposed to System II, it can be shown that the expected transmission rate for System I is at least as large as the expected rate for System II. 
	
	Note that in System II, there will be $\OO(m\cdot n^{1-\beta})$ users requesting non-empty contents, with high probability as $m$ grows large. Next, consider System III, where there is one combined cache of capacity $mk$ and $\OO(m\cdot n^{1-\beta})$ users, all with access to the combined cache, and each requesting a non-empty content with uniform probability. It is easy to see that the expected rate for System II is at least as large as the expected rate for System III.
	
	Finally, consider System IV, there is one combined cache of size $mk$ and $\OO(m\cdot n^{1-\beta})$ users, all with access to the combined cache, and each requesting an arbitrary non-empty content. It follows from \cite{niesen2014coded}[Claim 1] that the expected rate for System III is within a constant multiplicative factor of the worst-case rate in System IV, maximized over all feasible request vectors. 
	
	Combining the observations above, we have that the expected transmission rate for System I ($R^*_{\text{Zipf}}$) is at least as large as the worst-case rate in System IV ($R^*_{\text{Zipf, IV}}$), up to a constant multiplicative gap.  We now evaluate a lower bound on the worst-case rate for System IV. By considering multiple disjoint request vectors, each with unique requests made by every user, we have the following cut-set lower bound \cite{maddah2014fundamental}:
	\begin{eqnarray*}
		&& \dfrac{n}{\OO(m\cdot n^{1-\beta})} \EE[R^*_{\text{Zipf, IV}}]+ mk \ge n, \\ 
		\Rightarrow && \EE[R^*_{\text{Zipf, IV}}] \ge \Omega\left(\frac{(n - mk)m}{n^{\beta}}\right). 
	\end{eqnarray*}
\end{IEEEproof} 

\subsection{Proof of Theorem \ref{theorem:setting_C_converse}}

\begin{IEEEproof}[Proof of Theorem \ref{theorem:setting_C_converse}]
	Recall that each cache stores only one content. Let $y_{i,j}$ be equal to one if Content $i$ is stored on at least $j$ caches and zero otherwise. The expected number of caches used to serve requests for Content $i$ is upper bounded by $\sum_{j=1}^{\tilde{m}} y_{i,j} v_{i,j}$ where, $v_{i,j} = \sum_{k=j}^{\tilde{m}}p_i^{k} (1-p_{i})^{\tilde{m}-k}$ is the probability that Content $i$ is requested at least $j$ times.
	
	Therefore, the expected number of caches matched to serve user requests is upper bounded by $\text{O}_c$, where,	
	\begin{eqnarray*}
		\text{O}_c &=& \max \sum_{i=1}^n \sum_{j=1}^{\tilde{m}} y_{i,j} v_{i,j} \\
		&& s.t. \sum_{i=1}^n \sum_{j=1}^{\tilde{m}} y_{i,j} = m, \\
		&& y_{i.j} = \{0,1 \},\text{ } \forall i,j.
	\end{eqnarray*}	
	Since we have $\tilde{m}$ requests arriving in each time-slot, the expected transmission rate from the central server is lower bounded by $\tilde{m} -\text{O}_c$.
\end{IEEEproof}

\section{Related Work}
\label{section:related_works}

The problem of content replication for distributed caches has been widely studied. In \cite{SGSS14, SGSS14_2, LLM12,XT13,LLM13,Whitt07} the focus is on the setting where each request can be matched to any cache and the central server communicates with each user via an independent unicast transmission. In \cite{SGSS14, SGSS14_2} content popularity is unknown and time-varying, while \cite{LLM12,XT13,LLM13,Whitt07} study the setting where content popularity is either known or time-invariant or both. Unlike the Knapsack Storage policy, the proposed content replication policies in all these works store the more popular contents on a larger number of caches. The setting where each user is pre-matched to a server and the central server communicates with the users via an error free broadcast link has been studied in \cite{maddah2014fundamental, maddah2013decentralized, pedarsani2014online, niesen2014coded, hachem2014multi, zhang2015coded }. The idea of coded caching was introduced in \cite{maddah2014fundamental, maddah2013decentralized}. The key result in \cite{maddah2014fundamental, maddah2013decentralized, pedarsani2014online, niesen2014coded, hachem2014multi, zhang2015coded } is that unlike our setting, coded caching is necessary for optimal performance. Variants of the two settings discussed above have been studied in \cite{shanmugam2013femtocaching, borst2010distributed}.

\section{Conclusions}

The setting considered in this paper is motivated by large-scale distributed content delivery networks, used by Video on Demand services like Netflix and Youtube which have large content catalogs and serve a large number of users. 

We draw a parallel between the caching problem and the Knapsack problem and use it to design a caching and request routing policy called \emph{Knapsack Storage + Match Least Popular}. Surprisingly, our caching policy doesn't always cache the popular contents, and yet is order-optimal if content popularity follows the Zipf's law.
We also conclude that for the setting we consider, close to optimal performance can be achieved without using coded caching.


\bibliographystyle{IEEEtran}
\bibliography{myref2}


\appendix

In this section, we prove some of the results discussed in Section \ref{section:main_results} and Section \ref{section:proofs}.


\begin{IEEEproof} [Proof of Lemma \ref{lemma:Zipf_d}]	
	Since content popularity follows the Zipf distribution with Zipf parameter $\beta>1$,
	
	\begin{enumerate} 
		\item[(a)] For all contents less popular than Content $n_1$, 
		\begin{eqnarray*}
			p_{i} \leq \dfrac{p_1 (\log m)^2}{m}.
		\end{eqnarray*}
		Therefore, by the Chernoff bound (Lemma \ref{lemma:chernoff}), we have that, for $n_1 < i \leq n_2$,
		\begin{eqnarray*}
			\PP\big(d_i > 2p_1 (\log m)^2 \big) = \OO(e^{-(\log m)^{2}}).
		\end{eqnarray*}
		\item[(b)] For $i \leq n_1$, by the Chernoff bound (Lemma \ref{lemma:chernoff}), 
		\begin{eqnarray*}
			\PP\bigg(d_i > \bigg(1+\dfrac{p_1}{4}\bigg) mp_i \bigg) = \OO(e^{-mp_i}). 
		\end{eqnarray*}
		For $i \leq n_1$, $mp_i = \Omega((\log m)^{2})$, therefore, 
		\begin{eqnarray*}
			\PP\bigg(d_i > \bigg(1+\dfrac{p_1}{4}\bigg) mp_i \bigg) = \OO(e^{-(\log m)^{2}}).
		\end{eqnarray*}
	\end{enumerate}
	\noindent Therefore, by the union bound over all contents, we have that, 
	$\PP(E_1) = 1-\OO(n e^{-(\log m)^{2}}).$
\end{IEEEproof}

\begin{IEEEproof}[Proof of Lemma \ref{lemma:serve_all_requests}]
	Since the Match Least Popular policy matches requests to caches starting from the least popular contents, we first focus on requests for contents less popular than Content $n_2$. Since content popularity follows the Zipf distribution with Zipf parameter $\beta>1$, for $i > n_2$,
	\begin{eqnarray*}
		p_{n_2} < \dfrac{p_1 }{m(\log m)^{2}}.
	\end{eqnarray*}
	Each of these (ranked lower than $n_2$) contents is stored at most once across all caches. Therefore, under the Match Least Popular policy, a request for Content $i$ for $i > n_2$ will remain unmatched only if the cache storing that content is matched to another request for Content $i$ for some $i > n_2$. Since each cache stores at most $\tilde{k}$ contents, the cumulative popularity of all contents less popular than Content $n_2$ stored on a cache is $< \tilde{k}p_{n_2}$. Each unmatched request a Content $i$ for $i > n_2$ corresponds to the event that there are at least two requests for the $\tilde{k}$ contents less popular than Content $n_2$ stored on a cache. Therefore, by the Chernoff bound (Lemma \ref{lemma:chernoff}), the probability that a particular request for a Content $i$ for $i > n_2$ remains unmatched is $\leq \exp\bigg({-\dfrac{(\log m)^{2}}{\tilde{k}p_1}\bigg)}$. By the union bound, the probability that at least one request for Content $i \in R$ such that $i >n_2$ is not matched by the Match Least Popular policy is $\leq m \exp\bigg({-\dfrac{(\log m)^{2}}{\tilde{k}p_1}\bigg)}$.
	
	Next, we focus on contents ranked between $2$ and $n_1$. Note that, if the Knapsack Storage policy decides to store Content $i$, it stores it on $w_i$ caches. 
	\begin{eqnarray*}
		\sum_{i=2}^{n_2} x_i w_i &\leq& \sum_{i=2}^{n_1} \bigg\lceil \bigg(1 + \dfrac{p_1}{2}\bigg) mp_i \bigg\rceil + \sum_{i=n_1+1}^{n_2} \lceil 4 p_1 (\log n)^2 \rceil \\
		&\leq& \bigg(1 + \dfrac{p_1}{2}\bigg)  m(1-p_1)\\ 
		&& +  (4 p_1 (\log n)^2 + 1) m^{1/\beta}(\log m)^{2/\beta} \\
		&\leq& m.
	\end{eqnarray*}
	
	Therefore, if contents are stored according to Knapsack Storage: Part 2, each cache stores at most one content with index $i$ such that $2 \leq i \leq n_2$. 
	
	We now focus on requests for contents ranked between $2$ and $n_2$. Let $D_i$ be the set of caches storing Content $i$ for $2 \leq i \leq n_2$. Let $E_{3,j}$ be the event that cache $j \in D_i$ is matched to a request for Content $i$ for $i > n_2$. Since each cache stores at most $\tilde{k}$ contents, a cache is used to serve a request for Content $i$ for $i > n_2$ only if at least one of the $\leq \tilde{k}$ contents ranked lower than $n_2$ stored on the cache are requested at least once. Therefore,
	\begin{eqnarray*}
		\PP(E_{3,j}) \leq 1-\bigg(1- \dfrac{kp_1}{m(\log m)^{2}}\bigg)^m.
	\end{eqnarray*}
	For a given constant $\delta < 1$, $\exists m(\delta)$ such that for $m \geq m(\delta)$, $\PP(E_{3,j}) \leq \delta$. 
	
	For each Content $i$ for $n_1 < i \leq n_2$, let $E_{4,i}$ be the event that more than $2p_1(\log n)^2$ of the  $\lceil 4p_1(\log n)^2 \rceil$ caches in $D_i$ are matched to requests for Content $i$ for $i \geq n_2$. By the Chernoff bound for negatively associated random variables \cite{dubhashi1996balls}, 
	$
	\PP(E_{4,i}) = \OO\big(e^{-(\log n)^2}\big).
	$
	
	From Lemma \ref{lemma:Zipf_d}, we know that with probability $\geq 1-\OO\big(ne^{-(\log n)^2}\big)$, there are less than $2p_1(\log n)^2$ requests for each Content $i$ for $n_1 < i \leq n_2$. Therefore, with probability $\geq 1-\OO\big(ne^{-(\log n)^2}\big)$, all requests for contents in $R$ ranked between $n_1$ and $n_2$ are matched to caches by Match Least Popular. 
	
	For each Content $i$ for $2 \leq i \leq n_1$, let $E_{4,i}$ be the event that more than $\frac{p_1}{4}m p_1$ of the $w_i$ caches in $D_i$ are matched to requests for Content $i$ for $i \geq n_2$. By the Chernoff bound for negatively associated random variables \cite{dubhashi1996balls}, 
	$
	\PP(E_{4,i}) = \OO\big(e^{-(\log n)^2}\big).
	$
	
	From Lemma \ref{lemma:Zipf_d}, we know that with probability $\geq 1-\OO\big(ne^{-(\log n)^2}\big)$, there are less than $\big(1+\frac{p_1}{4}\big) mp_i$ requests for each Content $i$ for $2 \leq i \leq n_1$. Therefore, with probability $\geq 1-\OO\big(ne^{-(\log n)^2}\big)$, all requests for contents in $R$ ranked between $2$ and $n_1$ are matched to caches by Match Least Popular. 
	
	We now focus on the requests for Content 1. Recall that if the Knapsack Storage policy decides to cache Content 1, it is stored on all $m$ caches. Since the total number of requests in a batch is $m$, even if all requests for contents ranked lower than 1 are matched to caches, the remaining caches can be used to serve all the requests for Content 1. 
\end{IEEEproof}

\noindent \textbf{Proof of Corollary \ref{corollary:proportional_storage}}	
\begin{lemma}
	\label{lemma_G1}
	Let content popularity follow the Zipf distribution with Zipf parameter $\beta>1$. In a given time-slot, let $d_i$ be the number of requests for Content $i$. Let $G_1$ be the event that
	$$d_i \leq mp_i + \sqrt{mp_i}\log n \text{ for } 1 \leq i < \bigg(\frac{m}{\log m}\bigg)^{1/\beta}.$$
	Then we have that, 
	$$\PP(G_1) = 1-\OO(n e^{-(\log m)^{2}}).$$
\end{lemma}
\begin{IEEEproof}
	The number of requests for Content $i$ is a Binomial random variable with mean $mp_i$. Therefore, by the Chernoff bound (Lemma \ref{lemma:chernoff}),
	$$\PP(d_i \geq mp_i + \sqrt{mp_i}\log n) \leq e^{-(\log m)^2/3}.$$
	By the union bound, we have that, 
	$$\PP(G_1) = 1-\OO(n e^{-(\log m)^{2}}).$$
\end{IEEEproof}

\begin{lemma}
	\label{lemma_G2}
	Let content popularity follow the Zipf distribution with Zipf parameter $\beta>1$. In a given time-slot, let $d_i$ be the number of requests for Content $i$. Let $G_2$ be the event that
	$$\sum_{i = (\frac{m}{\log m})^{1/\beta} + 1}^n d_i = \OO(m^{1/\beta} \log m)$$
	Then we have that, 
	$$\PP(G_2) = 1-\OO(e^{-(\log m)^{2}}).$$
\end{lemma}
\begin{IEEEproof}
	The cumulative mass of all contents from $i = (\frac{m}{\log m})^{1/\beta} + 1$ to $n$ is
	$$\sum_{i=(\frac{m}{\log m})^{1/\beta} + 1}^n \frac{p_1}{i^\beta} = \OO\bigg( \bigg(\frac{\log m}{m}\bigg)^{\frac{\beta-1}{\beta}}\bigg).$$
	The total number of arrivals for all contents ranked lower than $(\frac{m}{\log m})^{1/\beta}$ in a time-slot is a Binomial random variable with mean $m^{1/\beta} (\log m)^{\frac{\beta-1}{\beta}}.$
	Therefore, 
	$$\PP\bigg(\sum_{i = (\frac{m}{\log m})^{1/\beta} + 1}^n d_i = \OO(m^{1/\beta} \log m)\bigg) \leq e^{-(\log m)^2/3}.$$
	Thus, $\PP(G_2) = 1-\OO(e^{-(\log m)^{2}}).$
\end{IEEEproof}
\begin{IEEEproof}[of Corollary \ref{corollary:proportional_storage}]
	Consider a policy called the Proportional Storage policy which stores at most one content per cache with Content $i$ stored on $c_i$ caches, such that,
	\begin{eqnarray*}
		c_i &=& \begin{cases}
			\lceil mp_i \rceil, & \text{for }1 \leq i \leq \bigg(\dfrac{m}{\log m}\bigg)^{1/\beta}, \\
			0, & \text{otherwise. }  
		\end{cases} 
	\end{eqnarray*}
	The quantity $\sum_{i=1}^n c_i$ can be upper bounded by a suitable integral to show that $\sum_{i=1}^n c_i \leq m$, i.e., Proportional Storage is a feasible storage policy. We skip the details due to lack of space. 
	
	Recall that $d_i$ is the number of requests for Content $i$ in a given time-slot. Therefore, under the Proportional Storage policy, the total number of transmissions from the central server is $$\sum_{i=1}^{(\frac{m}{\log m})^{1/\beta} + 1} (d_i - c_i) + \sum_{i = (\frac{m}{\log m})^{1/\beta} + 1}^n d_i.$$
	
	The rest of the proof is conditioned on events $G_1$ and $G_2$ as defined in Lemma \ref{lemma_G1} and \ref{lemma_G2}. Conditioned on $G_1 \cap G_2$, the expected number of transmissions from the central server is upper bounded by
	$$\bigg( \sum_{i=1}^{(\frac{m}{\log m})^{1/\beta} + 1} \sqrt{mp_i}\log n \bigg) + \OO(m^{1/\beta} \log m) = \OO(m^{1/\beta} \log m). $$
	
	Therefore, the expected number of transmissions from the central server for the Proportional Storage policy is upper bounded by
	\begin{eqnarray*}
		&&	\OO(m^{1/\beta} \log m) \PP(G_1 \cap G_2) + m (1-\PP(G_1 \cap G_2)) \\
		&=& \OO(m^{1/\beta} \log m) (1-\OO(n e^{-(\log m)^{2}})) + m \times \OO(n e^{-(\log m)^{2}})\\
		&\leq& \OO(m^{1/\beta} \log m) + \OO(m^2 e^{-(\log m)^{2}})\\
		&=& \OO(m^{1/\beta} \log m).
	\end{eqnarray*}
	
	By definition, this is also an upper bound on the expected number of transmissions from the central server for the optimal policy. 
\end{IEEEproof}

\noindent \textbf{Proof of Corollary \ref{corollary:knapsack_setting_C}} 


\begin{IEEEproof}[Proof of Corollary \ref{corollary:knapsack_setting_C}]	
	We consider the two cases separately. 
	\begin{enumerate}
		\item If $\tilde{k} = 1$, by the definition of the Knapsack Storage policy (\ref{fig:knapsack_storage_Part 1}), Content 1 is not stored on any of the $m$ caches. Given this, under Setting C, each request for Content 1 leads to a transmission from the central server. The expected number of requests for Content 1 in a time-slot is $mp_1 = \Theta(1)$, and therefore, the result follows. 
		\item If $\tilde{k} \geq \lceil c \rceil+2$, from the proof Theorem \ref{theorem:performance:our_policy}, we have that, with high probability, the Knapsack Storage + Match Least Popular policy serves all requests via the caches, i.e., no requests are served using the central server. Therefore, even for Setting C, the expected tranmission rate is $\oo(1)$. 
	\end{enumerate}
\end{IEEEproof}

\end{document}